\documentclass[usenatbib,usegraphicx]{mn2e}

\usepackage{times}

\usepackage{amssymb}

\newcommand{\re}{\ensuremath{R_\mathrm{e}}~}
\newcommand{\rs}[1]{\ensuremath{_\mathrm{#1}}}

\title[Simulations of the formation and evolution of isolated dwarf
galaxies]{Simulations of the formation and evolution of isolated dwarf
galaxies} \author[S. Valcke et al.]{S. Valcke$^{1}$\thanks{Doctoral
Fellow of the Fund for Scientific Research -- Flanders, Belgium
(FWO). E-mail: Sander.Valcke@UGent.be}, S. De
Rijcke$^{1}$\thanks{Postdoctoral Fellow of the Fund for Scientific
Research -- Flanders, Belgium (FWO). E-mail: Sven.Derijcke@UGent.be}
\& H. Dejonghe$^1$\\ $^{1}$Sterrenkundig Observatorium, Ghent
University, Krijgslaan 281, S9, 9000 Gent, Belgium}
\begin{document}

\date{Accepted . Received ; in original form }

\pagerange{\pageref{firstpage}--\pageref{lastpage}} \pubyear{2008}

\maketitle

\label{firstpage}

\begin{abstract}
We present new fully self-consistent models of the formation and
evolution of isolated dwarf galaxies. We have used the publicly
available N-body/SPH code \mbox{HYDRA}, to which we have added a
set of star formation criteria, and prescriptions for
chemical enrichment (taking into account contributions from both SNIa
and SNII), supernova feedback, and gas cooling. We extensively tested
the soundness of these prescriptions and the numerical convergence of
the models. The models follow the evolution of an initially
homogeneous gas cloud collapsing in a pre-existing dark-matter
halo. These simplified initial conditions are supported by the
merger trees of isolated dwarf galaxies extracted from the
milli-Millennium Simulation. The star-formation histories of the
model galaxies exhibit burst-like behaviour. These bursts are a
consequence of the blow-out and subsequent in-fall of gas. The amount
of gas that leaves the galaxy for good is found to be small, in
absolute numbers, ranging between $3\times10^7\ M_{\sun}$ and $6
\times 10^7\ M_{\sun}$. For the least massive models, however, this is
over 80 per cent of their initial gas mass. The local fluctuations in
gas density are strong enough to trigger star-bursts in the massive
models, or to inhibit anything more than small residual star formation
for the less massive models. Between these star-bursts there can be
time intervals of several Gyrs.

The models' surface brightness profiles are well fitted by S\'ersic
profiles and the correlations between the models' S\'ersic parameters
and luminosity agree with the observations. We have also compared
model predictions for the half-light radius \re, central velocity
dispersion $\sigma\rs{c}$, broad band colour $B-V$, metallicity
$[Z/Z_{\sun}]$ versus luminosity relations and for the location
relative to the fundamental plane with the available data. The
properties of the model dwarf galaxies agree quite well with those of
observed dwarf galaxies. However, the properties of the most massive
models deviate from those of observed galaxies. This most likely
signals that galaxy mergers are starting to affect the galaxies'
star-formation histories in this mass regime ($M \gtrsim 10^9\
M_{\sun}$).

We found that a good way to assess the soundness of models is provided
by the combination of \re and $\sigma\rs{c}$. The demand that these
are reproduced simultaneously places a stringent constraint on the
spatial distribution of star formation and on the shape and extent of
the dark matter halo relative to that of the stars.
\end{abstract}

\section{Introduction}
Dwarf galaxies (DGs), galaxies with blue absolute magnitude
$-14>M_\mathrm{B}>-18$ mag, are the most numerous type of galaxies in the
nearby universe \citep{ferguson1994}. Besides their number there are
various reasons why DGs are very important in modern astronomy, most
notably their cosmological relevance.

In the Lambda cold dark matter ($\Lambda$CDM) cosmological model,
present day galaxies are the product of a series of mergers. The early
universe consisted of a uniform density, superimposed with small
Gaussian fluctuations. As their amplitude increases monotonically
toward smaller scales, they give, after inflation, rise to a myriad of
small virialized dark matter (DM) haloes. These small haloes collapse first
under influence of their own gravity, followed by the collapse of
haloes with larger masses, swallowing the smaller haloes. We thus have
hierarchical clustering, where massive galaxies, such as our own Milky
Way, have formed through the merger of DGs. Dwarf galaxies are
therefore an important link in the chain of hierarchical
clustering. Moreover due to their relatively low total mass
($\mathcal{O}(10^7-10^9\ M_{\sun}$)) they are quite sensitive to the
effects of gravitational interactions, star formation (SF) and
supernova (SN) explosions. On the one hand DGs are therefore the ideal
objects to study when researching SF and SNe in a galaxy environment,
and dynamical evolution of galaxies in general. Understanding the
birth and life of DGs on the other hand could prove crucial when
trying to understand the formation and evolution of more massive
stellar systems. And even more: cosmological simulations of galaxy
formation exhibit very strong dependence on the way star formation,
feedback (FB) through supernovae and stellar winds (SW) are modelled
within stellar systems
\citep[see e.g.][]{white1991,kauffmann1999}. Therefore realistic
simulations of single dwarf galaxies can supply information on how to
implement star formation and feedback processes in large cosmological
simulations \citep[e.g.][]{springel2005}, where a fully consistent
implementation of SF and FB should be possible for moderate simulation
volumes.

From the theoretical point of view various approaches for the study of
DGs are possible, most notably there are semi-analytical models
(SAMs), consisting of a limited set of analytical equations solved
numerically, and fully numerical simulations, Lagrangian
(e.g. N--body/SPH) or Eulerian (grid codes). SAMs have the advantage of
computational efficiency, due to their speed it is possible to scan a
large part of the parameter space in a reasonable period of time. In
numerical simulations however a fully self consistent implementation
of SF, FB, chemical enrichment, gas cooling, ... is possible, at the
price of high computational effort.

\citet{dekel1986} were the first to model Dwarf Galaxies (DGs),
emphasising the impact of supernova-driven winds on the formation of
dwarfs. \citet{yoshii1987} found good agreement between observational
properties of DGs and those predicted by their SAM. These authors
calculated the properties of DGs (and globular clusters), using a
model that takes into account the dynamical response to a
supernova-driven wind. An evolutionary population synthesis method is
employed to bring the models into the observational domain.  According
to these models, spheroidal galaxies are a one-parameter family of
birth mass (initial total mass $M_\mathrm{t,i}$). These models were
later extended in \citet{naga2004}, who used a Monte Carlo technique
to construct the merger tree leading up to the formation of a
DG. \citet{naga2005} incorporated N--body simulations instead of a
Monte Carlo method to construct the merger tree. They found good
agreement between galaxies produced by their SAM, ranging from dwarf
elliptical to elliptical galaxies, and observations. SAMs are also
used to investigate certain processes involved in the birth and life
of dwarf galaxies rather than trying to reproduce dwarf galaxies,
e.g. the model of \citet{burkert1997}, investigating the role of SNIa
in delaying star formation and of \citet{ferrara2000} who investigated
the role of stellar feedback and dark matter.

Although plenty of numerical research has already been carried out
concerning the formation and evolution of massive elliptical galaxies
(EG) \citep[to name a few:][]{katz1991, kawata2001, chiosi2002,
kawata2003, kobayashi2005, merlin2006}, starting from various
cosmological initial conditions, only few people have tried directly
modelling DGs with fully numerical simulations. One problem is that
DGs are too small to be adequately resolved in large cosmological
simulations. A first fully numerical study of dwarfs was performed by
\citet{mori1997,mori1999}, who explored the role of gas cooling in
dwarf galaxy formation. \citet{chiosi2002} used an N--body/SPH code to
simulate the formation of galaxies, including dwarf spheroidal, dwarf
elliptical and normal elliptical galaxies. They started from
cosmologically inspired initial conditions. Assuming that star
formation starts after an initial merging period, the simulations
start with the collapse of an initially homogeneous gas cloud in a
dark matter halo \citep{white1978}. Observable properties of DGs were
broadly reproduced by their models. More recently, \citet{stinson2007}
conducted simulations of dwarf spheroidal galaxies (dSph). They used
initial conditions similar to those of \citet{chiosi2002}, except that
a degree of rotation was added to the gas. Like \citet{chiosi2002},
these authors find burst-like star formation. The work of
\citet{stinson2007} is based on the work by \citet{stinson2006}, who
investigated the influence of star formation criteria on the formation
of a galaxy, as well as the dependence of the processes involved on
the number of particles. \citet{mashchenko2007} performed fully
cosmological simulations of the formation of a few dwarf galaxies,
focusing on the gravitational heating of the dark matter, induced by
gas flows. They argue that this heating can effectively convert a
cusped dark matter halo into a DM halo with a central core. Other
studies of dSphs include those of \citet{read2005}, who looked at the
effect of baryonic mass loss from a two component galaxy (dark and
baryonic matter) and \citet{read2006}, who used cosmological
simulations to investigate the smallest baryonic building blocks for
galaxy formation. \citet{mashchenko2005} employed a simple model for
the formation of dwarf spheroidal galaxies. A single star-burst was
imposed on a gas distribution, employing a density cutoff, after which
the remaining gas was removed. They find reasonable agreement with
data from local group dwarf spheroidal galaxies.

A study of dwarf galaxies in a cosmological context was performed by
\citet{scannapieco2001}, investigating the effect of gas outflows on
the formation of dwarf galaxies. Other studies using simulations, not
directly aimed at reproducing dwarf galaxies, include those of
\citet{maclow1999}, who used a two-dimensional grid code to model
blowout/blow-away of gas in dwarf galaxies, and more recently
\citet{marcolini2006}, who used a three-dimensional grid code to study
the ISM in dwarf spheroidal galaxies by imposing a burst-like Star
formation history (SFH) and comparing the results to the Draco dSph.

In this paper we present the results of our models of dwarf galaxies,
with initial conditions taken from the $\Lambda$CDM formalism. As we
construct spheroidal, non-rotating galaxies, we expect them to
resemble dwarf spheroidal and dwarf elliptical galaxies. We took great
care to avoid the introduction of a direct dependence on the number of
particles or on the resolution into the formalism. We also investigate
the role of the different star formation criteria in detail. We then
compare our model data to observations using the central velocity
dispersion, the half-light radius, $B-V$ broad band colours, total
metallicity and the fundamental plane.

The outline of this paper is as follows. In \S~\ref{sec:code}, we
summarise the most important features of the modelling code. The
initial conditions of our models are discussed in
\S~\ref{sec:incond}. In \S~\ref{sec:results}, we compare our models
with observational data. We end with a summary and our conclusions in
\S~\ref{sec:sum}.

\section{The Code}

\label{sec:code}
We constructed our models using the N--body/SPH code \mbox{HYDRA}
\citep{couch1995,thacker2000}. In an N--body/SPH code the gravitational
equations are solved with an N--body integrator. Hydrodynamical forces
are applied through a Smoothed Particle Hydrodynamics (SPH) formalism
\citep{lucy1977,gingold1977}. In this formalism the density at the
position of the $i$-th (gas) particle is:
\begin{equation}
\rho_{\mathrm{g},i}=\sum_{j\neq i} m_j W(r_{ij},h_i),
\label{rhosph1}
\end{equation}
where $r_{ij}=|\vec{r}_i-\vec{r}_j|$, $h_i$ is the smoothing length of
the $i$-th particle and $W$ is the smoothing kernel. The use of the
SPH formalism lies in the fact that derivatives of
e.g. $\rho_\mathrm{g}$ can be expressed as derivatives of the
smoothing kernel $W$. \mbox{HYDRA} was modified by us to include star
formation, feedback, chemical enrichment, radiative cooling and
supernovae (type II and Ia).
\subsection{Star formation}
To implement star formation in our code we use a phenomenological,
Lagrangian approach. The first step is to formulate a set of star
formation criteria.
\subsubsection{Star Formation Criteria}
\label{sec:sfc}
We strive to select a set of Star-Formation Criteria (SFC) that allows
our simulations to mimic real-life star formation. A number of
implementations is compared in \citet{kay2002}, who find that
different implementations of SF give comparable results. We will use a
variant of the \citet{katz1996} criteria, where it is assumed that
star formation takes place in cool, dense, converging and
gravitationally unstable molecular clouds. A good review of these
criteria is given by \citet{stinson2006}. We select cool, dense and
converging clouds in a very straightforward manner: the temperature of
gas eligible for star formation must be lower than a certain critical
temperature $T\rs{c} = 15\,000\ \mathrm{K}$
\citep{stinson2006,stinson2007}. The density of a star forming gas
cloud must exceed a critical density $\rho\rs{g} = 2\times10^{-25}$
g cm-3, $n_{\mathrm{H}} \approx 0.1 \ \mathrm{cm-3}$
\citep{kawata2001,stinson2007}. Furthermore the local gas flow must be
converging: $\vec{\nabla}\cdot \vec{v} < 0$.

An important feature of a set of SFC is that they should be
independent of the number of particles. Not only is it a necessary
condition that simulations converge, we also want them to converge as
fast as possible to avoid extreme computation times. It is furthermore
impossible to keep the mass resolution fixed in simulations spanning a
wide mass-range (e.g. factor 100), because that would imply a
proportional variation of the number of particles
(e.g. $3\times10^4-3\times10^6$), resulting in drastic changes in and
large requirements for the number of particles. This reasoning rules
out the commonly used implementation of the Jeans-criterion for
dynamical instability~: $t_d \leq t_s$, with $t_d$ the dynamical time
and $t_s$ the sound crossing time, defined as $h/c_s = t_s$. Here $h$
is the SPH smoothing length, $c_s$ is the sound speed. Using $h$
introduces a direct relation to the number of particles. One should
moreover not attribute any physical meaning to it because it is an
artificial construct. Furthermore, as $t_d\sim1/\sqrt{\rho_g}$ and
$c_s\sim\sqrt{T}$, with $\rho_g$ the gas density and $T$ the gas
temperature, the Jeans-criterion is already contained in our
SFC. Therefore, instead of trying to find a meaningful length scale
for the gas, we did not include a separate criterion for gravitational
instability of the gas. Our SFC are thus:
\begin{eqnarray}
		\vec{\nabla}\cdot\vec{v} & \leq & 0 \label{sfcrit1}\\
		\rho_g & \geq & \rho_c = 0.1\ \mathrm{cm^{-3}}
		\label{sfcrit2}\\ T & \leq & T_c =
		15000\,\mathrm{K}. \label{sfcrit3}
\end{eqnarray}
Each time step the SFC are checked for each gas particle. When a gas
particle meets all three conditions it becomes eligible for star
formation. Although one should refrain from attributing particle
properties to the SPH gas particles because they are actually (only)
co-moving grid points, this course of action is justified when
interpreted as a particle number independent check of the SFC
throughout our gas cloud. As there are more grid points (particles)
where the gas density is higher, this check is performed more in
highly clustered regions, which makes perfect sense when considering
our SFC (eq. \ref{sfcrit1}--\ref{sfcrit3}). This does however not
imply a dependence of the particle number, because the newly formed
star particles have a mass proportional to their parent gas
particles. When e.g. doubling the number of gas particles (in a
certain area, or in the entire simulation), the mass attributed to
each gas particle is cut in half. As our SFC are independent of the
number of particles, two times as much gas particles will form stars,
leaving the total mass of newly formed stars invariant. The number of
gas particles in a certain area thus determines the average mass of
star particles spawned in that area and the resolution with which the
SFC are checked, it does not (or very lightly) influence the total
star mass generated in that area.
\subsubsection{Star Formation Law}
When we have selected those particles eligible for star formation, we
have to prescribe a recipe to turn them into stars. For our SF law we
adopt the Schmidt law of (1959)~:
\begin{equation}
\frac{\mathrm{d}\rho_\mathrm{s}}{\mathrm{d}t} 
= - \frac{\mathrm{d}\rho_\mathrm{g}}{\mathrm{d}t}
= c_*\frac{\rho_\mathrm{g}}{t_\mathrm{g}},
\label{schmidt1}
\end{equation}
where $\rho_\mathrm{s}$ and $\rho_\mathrm{g}$ are the density of stars
and gas respectively, $c_*$ is a dimensionless scaling
constant. $t_\mathrm{g}$ Is a characteristic time-scale for the gas,
we adopt $t_\mathrm{g}=t_\mathrm{d}$, where $t_\mathrm{d}$ is the
dynamical time. We define the dynamical time as:
\begin{equation}
t_\mathrm{d} = \frac{1}{\sqrt{4\pi G\rho_\mathrm{g}}},
\label{dyntime1}
\end{equation}
where G is the gravitational constant. Throughout our simulations we
set $c_*=1$. It is possible to constrain $c_*$ and $\rho\rs{c}$ by
fitting star formation rates to the Kennicut--Schmidt law
\citep{kenni1998}. \citet{stinson2006} found little influence of $c_*$
on the median SFR when going from 0.05 to 1 (their fig. 14). They
opted to use 0.05, we use the canonical value 1.

We argue that the use of
$\rho_\mathrm{tot}$ instead of $\rho_\mathrm{g}$, as argued in
\citet{buonomo2000}, in the definition of the dynamical time is less
justified. \citet{buonomo2000} argue that the use of
$\rho_\mathrm{tot}$ results from deriving the Jeans criterion for
instability \citep[see e.g.][]{binney1987} for a gas cloud embedded in
a dark halo. This is only true if the dark matter collapses along with
the gas but not if the dark matter merely acts as a background fluid
with, locally, constant density. We therefore use only the gas density
$\rho_\mathrm{g}$ in our definition of $t_\mathrm{d}$.

Integrating equation (\ref{schmidt1}) over the interval $\Delta t$
(assuming constant $t_d$), multiplying with a volume $V$ and dividing
by the initial gas mass $M_\mathrm{g,0}$ gives the probability that a
gas particle forms a star particle:
\begin{equation}
p^* = 1-\exp{\left(-\frac{\Delta t}{t_\mathrm{d}}\right)}
\label{schmidt2}
\end{equation}
This equation has the immediate advantage that it introduces a
quasi-independence of the number of time-steps into the formalism. If
we bridge a time-interval $\Delta t$ with one step, the $P$ to form a
star particle is: $P = 1-\exp(-\Delta t / t\rs{d})$. For small values
of $\Delta t/t\rs{d}$ we have: $P\approx \Delta t/t\rs{d}$. For the
total probability to form a star particle using 2 steps we then have:
$(1-\Delta t/(2t\rs{d}))\Delta t/(2t\rs{d})+\Delta
t/(2t\rs{d})(1-\Delta t/(2t\rs{d}))=\Delta t/t\rs{d}-2(\Delta
t/(2t\rs{d}))^2$. This is, to the first order in $\Delta t/t\rs{d}$,
equal to the single-step value. As $\Delta t/t\rs{d}$ is typically
0.01 (see further), the approximation is justified. Equation
(\ref{schmidt2}) is implemented using a Monte-Carlo procedure: a
random number $r$ between 0 and 1 is drawn, if $r<p^*$ the gas
particle gives birth to a star particle. The mass of a newly formed
star particle is fixed at $1/3$ of the mass of the parent gas
particle. Each gas particle is allowed to form 4 star particles after
which its remaining mass, metals, energy, ... are distributed among
neighbour gas particles, using the SPH kernel. \cite{kenni1998} finds
that the median rate of gas consumption is 30 per cent per $10^8\
\mathrm{yr}$. Calculating the average central dynamical time
$t_\mathrm{d,c}$ (equation (\ref{dyntime1}) for gas particles within a
radius of 0.5 kpc from the centre) we find a value $\mathcal{O}(10^8\
\mathrm{yr})$. The formula for the average mass of a gas particle
after $N$ equal time-steps, taking into account the efficiency of 1/3
for SF as well as equation (\ref{schmidt2}), is:
\begin{equation}
\bar{M}=M_0\left[ 1+\sum_{i=1}^N 
\left[\left(\frac{2}{3}\right)^{\min(i,4)}-1\right]p^*(1-p^*)^{N-i}C^i_N\right],
\label{mbar1}
\end{equation}
where $C^i_N=N!/(i!(N-i!))$, $p^*(1-p^*)^{N-i}C^i_N$ is the
probability that a star particle will form stars i times in N steps
and $\left(2/3\right)^i$ stems from the 1/3 efficiency of our SF. When
using equation (\ref{mbar1}) we assume that the gas particle satisfies
the 3 SFC for the duration of the N time-steps. We also assume that
$p^*$ is invariant during these time-steps. The limited number of star
formation episodes (4) is taken into account by using $\min(i,4)$
instead of $i$, which comes down to setting the efficiency in terms of
mass of more than 4 SF episodes equal to 4. This yields a negligible
correction. We can write equation (\ref{schmidt2}) as
$p^*\approx1-\exp(-1/N)$, with $N=t_\mathrm{d}/\Delta t$. As $t_d\
\mathcal{O}(10^8)$, and the time-steps in our simulations are
$\mathcal{O}(10^6)$, we have $N\ \mathcal{O} (100)$. Using this value
gives $\bar{M}\approx 0.72\ M_0$, so our factor $1/3$ models 30 per
cent efficiency rather well. This is of course a crude estimation, but
we preferred it over adding a parameter to our model, which would
multiply the required number of simulations by a further factor.
\subsection{Feedback}
There are two mechanisms through which stars can return metals to
their environment: stellar winds and supernovae. These can be modelled
given knowledge of the initial mass function (IMF), which determines
the distribution over mass within a ``star particle'' (single age
single metallicity particle or SSP). We have adopted the Salpeter IMF
in our simulations:
\begin{equation}
	\Phi(m)\mathrm{d}m = Am^{-(1+x)}\mathrm{d}m,
	\label{salpeter1}
\end{equation}
where $x=1.35$ and $A$ is fixed by the chosen
normalisation. $\Phi(m)(=\mathrm{d}N/\mathrm{d}m)$ is the probability
that a star with mass $m$ resides in the SSP. We normalise the total
probability to one:
$\int_{m_\mathrm{l}}^{m_\mathrm{u}}\Phi(m)\mathrm{d}m=1$. Using
$m_\mathrm{l}=0.1\ \mathrm{M}_{\sun}$ and $m_\mathrm{u}=60\
\mathrm{M}_{\sun}$ we have $A=0.06$. To calculate the average value of
a certain quantity $B$ of a SSP we then have:
\begin{equation}
	\left\langle B\right\rangle = \int_{m_\mathrm{B,l}}^{m_\mathrm{B,u}} B(m)\Phi(m)\mathrm{d}m \times
	\frac{M_\mathrm{SSP}}{\int_{m_\mathrm{l}}^{m_\mathrm{u}}m\Phi(m)\mathrm{d}m},
	\label{avform1}
\end{equation}
with $m_\mathrm{B,l}$ and $m_\mathrm{B,u}$ respectively the lower and
upper bound for the mass interval where $B$ applies, and
$M_\mathrm{SSP}$ the mass of the stellar particle. Application of
equation (\ref{avform1}) for energy feedback of SN II, where we assume
$B(m)=E_\mathrm{SNII}$ to be independent of mass, and with
$m_\mathrm{SNII,l}=8\ \mathrm{M}_{\sun}$, $m_\mathrm{SNII,u}=60\
\mathrm{M}_{\sun}$, gives:
\begin{eqnarray}
	E_\mathrm{tot,SNII} 
	&=& E_\mathrm{SNII}\int_{8\ \mathrm{M}_{\sun}}^{60\ \mathrm{M}_{\sun}} \Phi(m)\mathrm{d}m \nonumber\\
	&&\hspace{5 ex}\times
	\frac{M_\mathrm{SSP}}{\int_{0.1\ \mathrm{M}_{\sun}}^{60\ \mathrm{M}_{\sun}}m\Phi(m)\mathrm{d}m},
\label{sniiener1}
\end{eqnarray}
which gives us:
\begin{equation}
E_\mathrm{tot,SNII}=7.31 \times 10^{-3}M_\mathrm{SSP}E_\mathrm{SNII}\mathrm{M}_{\sun}^{-1}.
\label{sniiener2}
\end{equation}
For SWs the derivation is, apart from replacing $E_\mathrm{SNII}$ by
$E_\mathrm{SW}$, identical. Following \citet{thornton1998} we set
$E_\mathrm{SNII}=10^{51}\ \mathrm{erg}$, $E_\mathrm{SW}$ is set to
$=10^{50}\ \mathrm{erg}$. The actual energy returned to the ISM, for
SNe as well as for SW, is implemented as $\epsilon_\mathrm{FB}\times
E_\mathrm{tot}$, where $\epsilon_\mathrm{FB}$ was chosen to be 0.1,
following the results of \citet{thornton1998}. The returned mass
fraction $F_\mathrm{SNII}$ can be easily calculated by subtracting the
total mass that is not returned to the interstellar medium (ISM) by
exploding stars from the total mass of stars eligible to go
supernova. With the approximation that the mass remaining in dark
objects (black holes, neutron stars, white dwarfs) after a star goes
SN II is constant, $M_\mathrm{rem}\approx1.4\ \mathrm{M}_{\sun}$, we
have:
\begin{eqnarray}
	F_\mathrm{SNII} &=& 
	\frac{\int_{m_\mathrm{SNII,l}}^{m_\mathrm{SNII,u}}m\Phi(m)\mathrm{d}m
	- M_\mathrm{rem}\int_{m_\mathrm{SNII,l}}^{m_\mathrm{SNII,u}}\Phi(m)\mathrm{d}m}
		{\int_{m_\mathrm{l}}^{m_\mathrm{u}}m\Phi(m)\mathrm{d}m}\nonumber\\
	 &=& 0.112.
	\label{sniimafra1}
\end{eqnarray}
The yield of element $i$ by SN II is calculated as (equation (\ref{avform1})):
\begin{equation}
	M_i = 
	M_\mathrm{SSP}\frac{\int_{m_\mathrm{SNII,l}}^{m_\mathrm{SNII,u}}M_i(m)\Phi(m)\mathrm{d}m}
		{\int_{m_\mathrm{l}}^{m_\mathrm{u}}m\Phi(m)\mathrm{d}m}.
	\label{sniiyie1}
\end{equation}
For the metal yields of stars with mass $m$ we fitted cubic splines to
the data points tabulated by \citet{tsujimoto1995}. For SN Ia the
structure of the used formulae is analogous, the only difference being
the inclusion of a factor $A_\mathrm{SNIa}$ in every right hand side
of the above equations. $A_\mathrm{SNIa}$ represents the fraction of
stars in the mass range of SN Ia that actually go supernova, because
SN Ia are believed to take place after a period of Roche lobe overflow
in a binary star system with a white dwarf and a red giant
companion. We calculate $A_\mathrm{SNIa}$ as follows
($m_\mathrm{SNIa}$, the lower limit for SN Ia, is taken to be $3\
\mathrm{M}_{\sun}$, the upper limit for SN Ia is
$m_{\mathrm{SNII,l}}$):
\begin{equation}
	\frac{N_\mathrm{SNIa}}{N_\mathrm{SNII}}=0.15=A_\mathrm{SNIa}
	\frac{\int_{m_\mathrm{SNIa}}^{m_\mathrm{SNII,l}}\Phi(m)\mathrm{d}m}
		{\int_{m_\mathrm{SNII,l}}^{m_\mathrm{SNII,u}}\Phi(m)\mathrm{d}m},
	\label{sniabiga1}
\end{equation}
where 0.15 was found by \citet{tsujimoto1995} as the number ratio that
best reproduces the observed abundance pattern among heavy elements in
the solar neighbourhood. This gives us: $A_{\mathrm{SNIa}}=0.0508$
(compare to e.g. \citet{kawata2001} who use 0.04). We note that
because SN Ia take place when white dwarfs reach the Chandrasekhar
limit ($1.4\ \mathrm{M}_{\sun}$), equation (\ref{sniiyie1}) will be
simplified for SN Ia because $M_i(m)$ becomes $M_i$, independent of
mass.

Using $A_\mathrm{SNIa}$ and an equation analogous to equation
(\ref{sniimafra1}), setting $M_\mathrm{rem}=0$, we find:
$F_\mathrm{SNIa}=0.00502$. Using the analogon of equation
(\ref{sniiyie1}) for SN Ia, keeping in mind that the yields per
supernova are independent of the progenitor mass, we find:
$M_{i,\mathrm{SNIa}}=F_{i,SNIa}M_{\mathrm{SSP}}$ with $F_{i,SNIa} =
0.0011 m_{i,\mathrm{SNIa}}\mathrm{M}_{\sun}^{-1}$. We used the yields
given in \citet{trava2004} (their b20\_3d\_768 model).

For the time intervals for SNe and stellar wind we use a formula for
the main-sequence lifetime of a star with mass $m$ \citep{david1990}:
\begin{equation}
	\log{\tau(m)} = 10 - 3.42\log{m} + 0.88 \left(\log(m)\right)^2.
	\label{mstime1}
\end{equation}
Filling in the various lower- and upper bounds for the mass of SNe we
have: SW: $0 - 4.3\times10^7$ yr, SNe II:
$5.4\times10^6-4.3\times10^7$ yr, SNe Ia: $1.543\times10^9 -
1.87\times10^9$ yr. For SN Ia a net delay of 1.5 Gyr was applied,
following the result of \citet{yoshii1996}, who found 1.5 Gyr for the
mean lifetime of SNe Ia based on a chemical evolution model applied to
the solar neighbourhood. This time interval is the result of the time
needed for the accretion of mass by the WD to push its mass up to the
Chandrasekhar limit. For the distribution of feedback energy we used
the simplest prescription: a constant amount of feedback over the time
intervals.

The actual feedback uses the SPH smoothing kernel: if star particle
$i$ with smoothing length $h_s$ distributes an amount of energy $E$,
then a neighbouring gas particle $j$ receives an amount of energy
$\Delta E_j$:
\begin{equation}
   \Delta E_j = 
   \frac{m_j W(|\vec{r}_i-\vec{r}_j|,h_s)E}{\sum_{k=1}^Nm_kW(|\vec{r}_i-\vec{r}_k|,h_s)}.
\end{equation}
Metallicity-dependent cooling (with a maximum metallicity of 1
$Z_{\sun}$) is implemented using the tables given by
\citet{sutherland1993}, allowing our gas to cool to a minimum
temperature of $10^4$ K (cooling beyond this temperature is possible
however through adiabatic expansion). To compensate for the fact that
we are unable to resolve the hot, low density cavities in the gas,
originating from SN explosions, we implemented a period where
radiative cooling is switched off, only allowing adiabatic cooling
\citep{thacker2000fb}. To be precise a particle is not allowed to cool
radiatively during a certain time-step if it was heated by SNe during
that time-step.

As explained in \citet{bate1997}, a difference of the gravitational
softening ($\epsilon$) and SPH smoothing ($h$) can lead to artificial
convergence/divergence of SPH particles. We tested $\epsilon=30$ pc
and $\epsilon=60$ pc and found very comparable simulation results,
permitting us to use 60 pc. $h$ ranges from on average 120 pc (C02
model) to 60 pc (C09 model), so $\epsilon$ and $h$ are of comparable
size. As the global evolution of our models is determined by the
in-fall into the deep potential well, we do not have to be very
concerned about mismatching of $\epsilon$ and $h$ on that scale. And
as the resolution of our simulations is insufficient to capture the
collapse of small ($\mathcal{O}(30$ pc)) molecular clouds, we need not
be concerned about artificial convergence/divergence of SPH particles
on that scale.

Broad band colours of our models are calculated (with bi-linear
interpolation) using the models of \citet{vazdekis1996}, who provide
Mass/Luminosity values for SSPs according to metallicity and age.
\section{Initial Conditions}
\label{sec:incond}
As initial conditions, we use a relatively simple setup, assuming a
flat $\Lambda$-dominated cold dark matter cosmological model. We
further assume that in the process of dwarf galaxy formation, star
formation comes into play only after there has already been
substantial merging. Support for this assumption can be gleaned from
\citet{naga2004}, who find that SAMs with long star-formation
time-scales at high redshift provide the best agreement with the
observed properties of dwarf galaxies. Furthermore, we used the
freely available results from the milli-Millennium
Simulation\footnote{http://www.g-vo.org/Millennium}
\citep{springel2005} to assess the formation redshift of the halos
that at redshift $z=0$ fit the description of an isolated dwarf galaxy
halo. The dwarf galaxies modeled here are actually below the
resolution of the Millennium Simulation, which requires a minimum of
$N\rs{p}=20$ particles, or a mass of about $1.7\times10^{10} M_{\sun}
h^{-1}$, to properly identify a halo. This corresponds to the mass of
the C09 halo, our most massive model (see table \ref{tabm})).

We queried the milli-Millennium Simulation for the formation redshift
of halos that at $z=0$ would be associated with isolated dwarf
galaxies. We defined the formation redshift as the redshift at which a
halo achieved approximately $2/3$ of its final mass, allowing for a
factor of 1.75 of further mass growth through minor mergers. Halos
that undergo a major merger after their formation are explicitly
excluded from this tally. We cannot derive the formation redshifts of
halos with particle numbers $N\rs{p} \approx 20$ at $z=0$ because {\em
(i)} the closer the halo particle number gets to the lower limit of
halo detection, the less accurate we can determine the actual time of
birth of that halo and {\em (ii)} for such low particle numbers it is
impossible to construct a proper merger tree. Instead, we estimate the
formation redshifts of halos with particle numbers between 35 and 45
at $z=0$ by querying when they first encompassed at least 20
particles. In order to investigate whether the halo formation redshift
depends on galaxy mass, we also queried the milli-Millennium
Simulation for the redshift at which halos with particle numbers
$175<N\rs{p}<185$ at $z=0$ first encompassed 100 particles. This again
alows for a factor 1.75 of further mass growth through minor mergers.
\begin{figure}
\includegraphics[width=\hsize]{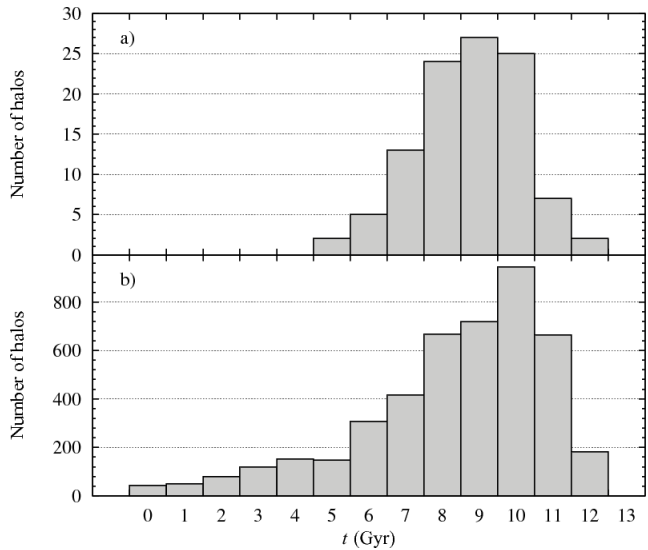}
\caption{The number of halos drawn from the Millenium Simulation with
particle numbers a) $175<N\rs{p}<185$ and b) $35<N\rs{p}<45$ at $z=0$,
binned as a function of how many gigayears ago these halos first
reached particles number above a) 100 and b) 20. Halos that undergo a
major merger after their formation are explicitly excluded from this
tally.}
\label{mil_hist}
\end{figure}

It is clear from Fig. \ref{mil_hist} that the peak of halo formation
is situated at about 10~Gyr ago, with the majority of halos having
formed between 8 and 11~Gyr ago. A closer inspection of both panels of
Fig. \ref{mil_hist} reveals that the formation rate of the least
massive halos, presented in panel b), peaks at $\approx 10$~Gyr
ago. This is about 1~Gyr earlier than the formation rate of their more
massive analogs, presented in panel a). Extrapolating towards the even
lower masses of the galaxies we modeled for the present paper, one
could expect the halo formation time to shift to even earlier times,
with the oldest halos being formed as early as 12~Gyr ago. We will,
therefore, assume that the dark-matter halos of the dwarf galaxies we
modeled formed at a redshift $z=4.3$ and neglect further growth of the
halo. Furthermore we neglect any environmental effects
(e.g. ram-pressure stripping) on the chemical and dynamical evolution
of the models. These environmental influences are not always
negligible (see e.g. \citet{penarrubia2007}). Thus, the models
describe dwarf galaxy formation as the isolated collapse of a gas
cloud in the potential well generated by a spherical dark matter halo
and the gas itself \citep[see e.g.][]{gao2007}.

We use the following cosmological parameters: $h=0.71$,
$\Omega_\mathrm{tot}=1,
\Omega_\mathrm{matter}=\Omega_\mathrm{m}=0.2383,
\Omega_\mathrm{DM}=0.1967$ \citep{spergel2007}, where $h$ is the
normalised Hubble constant. The mean density of the Universe as a
function of redshift $z$ and Hubble constant $H_0=100\,h$ is:
\begin{eqnarray}
   \rho_\mathrm{u}
   &=&\frac{3h^2100^2}{8\pi G}\left[\Omega_\mathrm{m}(1+z)^3+1-\Omega_\mathrm{m}\right]\nonumber\\
&\hspace{-3ex}=&\hspace{-3ex}
1.99\times10^{-29}h^2\left[\Omega_\mathrm{m}(1+z)^3+1-\Omega_\mathrm{m}\right]\,\mathrm{g}\,\mathrm{cm}^{-3}.
	\label{densuniv1}
\end{eqnarray}
We take the over-density of matter to $\rho_\mathrm{u}$ to be 5.55,
the value when the local flow detaches itself from the Hubble flow,
according to the Tolman model of a spherical over-density. $z$ at the
start of the simulations is taken to be 4.3, corresponding to
approximately 1.5 Gyr after the birth of the universe. The simulation
length is set to 10 Gyr, which is long enough to cover the most
important part of the star forming period of the simulated
galaxies. The density of matter at the start of the simulations is
then:
\begin{equation}
	\rho_{\mathrm{m}}=5.55\times\rho_u.
	\label{densbar1}
\end{equation}
When the total mass and its density are known we can fix the outer radius:
\begin{eqnarray}
	R_{5.55}(z)
	&=&\left(\frac{3}{4\pi}\right)^{1/3}\left(\frac{M_\mathrm{T}}{\rho_{\mathrm{m}}}\right)^{1/3}
	\nonumber\\
	&\hspace{-10ex}=&\hspace{-6ex}\frac{0.09617}{5.55^{1/3}}
\left[\frac{M_\mathrm{T}}{h^2\left(\Omega_\mathrm{m}(1+z)^3+1-\Omega_\mathrm{m}\right)}\right]^{1/3},
	\label{radbar1}
\end{eqnarray}
where $M_\mathrm{T}$ is expressed in $\mathrm{M}_{\sun}$ and
$R_{5.55}$ in kpc. At the start of the simulations the gas particles
are randomly distributed in a sphere with radius given by equation
(\ref{radbar1}). The gas particles are initially at rest, their
initial metallicity is set to $10^{-4}\ Z_{\sun}$ and their initial
temperature is $10^4$ K.

Several choices are possible for a DM profile. There is the NFW
profile, deduced from simulations in \citet{navarro1997}. However
observations as well as simulations seem to imply that the DM halo has
a central core rather than a cusp \citep[][and references
therein]{salucci1997,merritt2006,gentile2007}. Other previously used
profiles include the modified isothermal sphere \citep{binney1987} or
the King profile \citep[e.g.][]{mori1999}, see \citet{binney1987}. For
the DM halo we choose the Kuz'min Kutuzov (KK) profile as presented by
\citet{dejo1988}. The KK profile has several distinct advantages~: it
is, apart from a numerically well-behaved integral, analytical, and it
has a central core. Furthermore, \citet{sell1997} have carried out a
stability analysis of the KK halo, concluding that haloes with
flattening up to E7 are stable. The density of this model is given by:
%
%
\begin{eqnarray}
	\rho_\mathrm{KK}(R,z)&=&\frac{Mc^2}{4\pi}\frac{(a^2+c^2)R^2+2a^2z^2+}
		{(a^2c^2+c^2R^2+a^2z^2)^{3/2}}\nonumber\\
	&& \hspace*{-3em} \frac{2a^2c^2+a^4+3a^2\sqrt{a^2c^2+c^2R^2+a^2z^2}}{(R^2+z^2+a^2+2\sqrt{a^2c^2+c^2R^2+a^2z^2})^{3/2}},
\label{kuzkutrho1}
\end{eqnarray}
with $a$ and $c$ parameters controlling the length of the semi-major
and semi-minor axis respectively, $R$ is the distance in the
$xy$--plane and $z$ is the distance normal to this plane.

We generated the DM halo using a standard Monte Carlo sampling
technique. For each particle we first generate three coordinates
$R,\phi,z$, distributed as $\rho_\mathrm{KK}$ (equation
(\ref{kuzkutrho1})). Then $v_\mathrm{R},v_\mathrm{\phi}$ and
$v_\mathrm{z}$ are found using a standard acceptance-rejectance
technique using the distribution function of the KK model. The halo
density is set to zero outside the equipotential surface
$\Psi_\mathrm{KK}(R,z) = \Psi_\mathrm{KK}(R_\mathrm{cut},0)$, with
$R_\mathrm{cut}=25$ (units $(a+c)$). This large value ensures that the
effect of this cut-off on the halo's stability is negligible. We
explicitly checked the halo's stability by letting it evolve in
isolation for the same duration as the science simulations. 

Now that we have fixed the cosmological parameters (including the
ratio of DM mass to gas mass), we are left with 3 parameters
determining the model: $M_\mathrm{tot},a$ and $c$. $c$ determines the
amount of flattening we want for the halo, where $a=c$ gives a
spherical halo. As we restrict ourselves here to spherical systems we
set $c=a$. To eliminate a further parameter we set, analogous to
$R_{5.55}$ in equation (\ref{radbar1}),
\begin{equation}
	a = C \sqrt[3]{M_\mathrm{tot}}\ \frac{\textrm{kpc}}{\sqrt[3]{M_{\sun}}},
	\label{kka1}
\end{equation}
where $C$ is fixed by associating one value of $a$ with a value of
$M_\mathrm{tot}$. Several values of $C$ are explored with our
models. This leaves us with \emph{one parameter} to vary in a series
of simulations: $M_\mathrm{tot}$.
\section{Results}
\label{sec:results}
In the following sections we discuss the models. Information about
these models can be found in Tables \ref{tabnpart} (A models) and
\ref{tabm} (C and D models). The A models all have an identical setup,
apart from the number of gas particles. The number of DM particles is
set to $15\,000$. We use these A models to determine the minimum
number of gas particles that is required for the simulations to be
numerically converged. The C models represent the ``real'' models. We
compare in detail some structural parameters of the simulated C-model
galaxies with those of observed dwarf galaxies, in order to assess how
well the simulations reproduce reality. The B models are essentially a
larger version of the C models, being identical in setup, apart from a
smaller value of $C$ (equation (\ref{kka1}): $C = 0.000928$. We also
sampled the mass range differently: the total mass of the Bx model
does not necessarily correspond to the mass of the Cx model (B models:
0.25, 0.3, 0.65, 1, 3, 6 $\times 10^9\ M_{\sun}$). This however has no
influence on a comparison of the B and C models. The B models are
included only to illustrate the effect of varying $C$ on the
half-light radius and the central velocity dispersion of the model
galaxies (sections \ref{subsec:halfl} and \ref{subsec:vd}). They were
constructed using $15\,000$ dark matter particles. The two D models do
not obey equation (\ref{kka1}). They were constructed using an
arbitrary (high) value of $C$. The D models are included to show that
it is difficult to extend our C models beyond the mass where they
break down by fiddling with some parameters.

\subsection{Number of particles}
\begin{figure}
\includegraphics[width=\hsize]{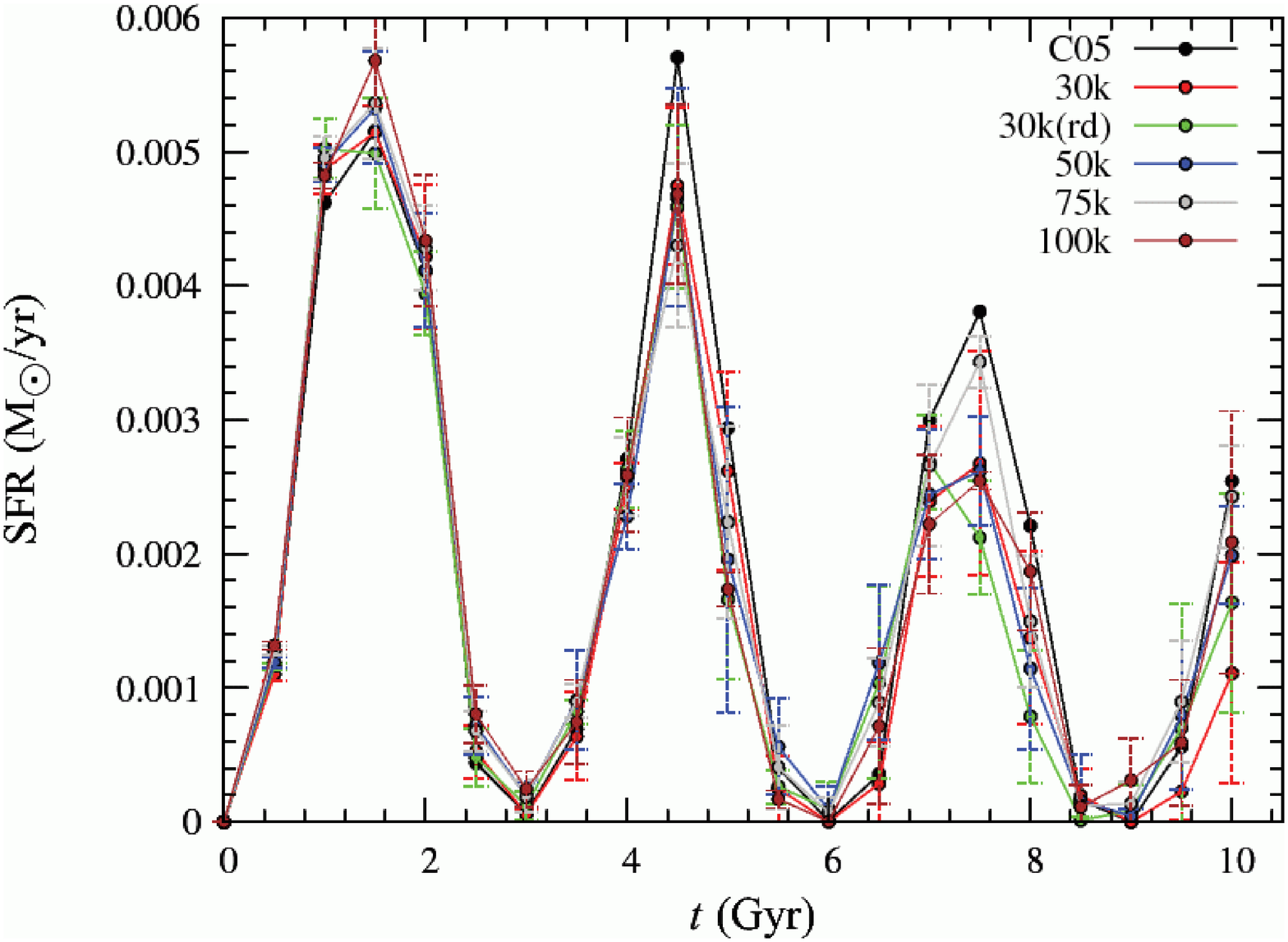}
\caption{Star Formation Rate (SFR, in M$_{\sun}$ yr$^{-1}$) as a
function of time (Gyr) for simulations with varying particle
numbers. See text and table \ref{tabnpart} for more details.}
\label{npart_sfr}
\end{figure}

\begin{figure*}
\includegraphics[width=\hsize]{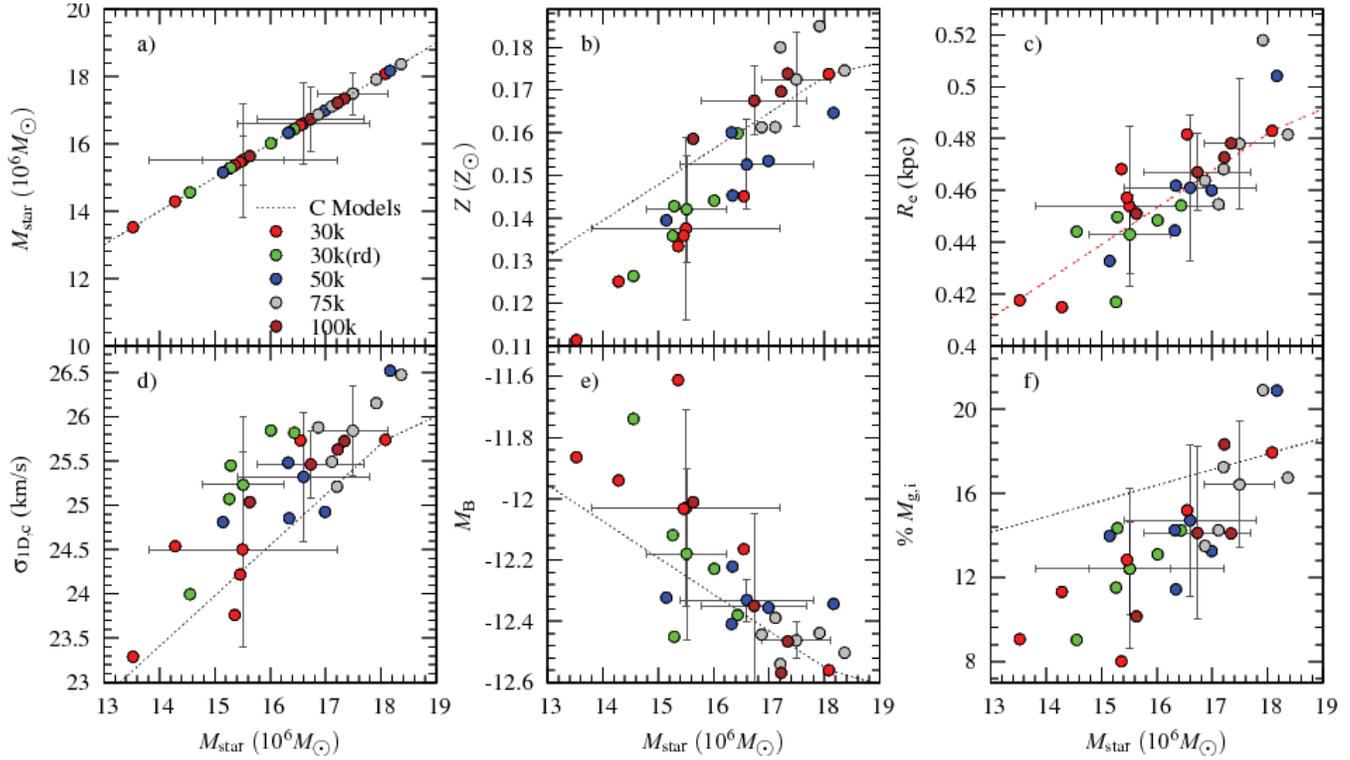}
\caption{Variation of model observables as a function of gas particle
number, shown for the models listed in table \ref{tabnpart}. The
dash-dotted line connects the points of the C models. The symbol at
the center of the error-bars indicates to which models these bars
belong (it is not a datapoint).}
\label{npart_statistics}
\end{figure*}

To be able to compare simulations covering a wide range of initial
masses, the simulations have to converge numerically. At a certain
point a further increase of the number of particles should make no (or
very little) difference. For this reason we took special care in
selecting the SFC (equations \ref{sfcrit1}--\ref{sfcrit3}). Results of
a first set of simulations for a range of particle numbers (A models)
are shown given in
Figs. \ref{npart_sfr}--\ref{npart_statistics}. These models all use
the parameters of the C05 model (see table \ref{tabm}).
\begin{table}
\caption{Information concerning the simulations with varying particle
numbers. See text for more details.}
\label{tabnpart}
\centering
\begin{tabular}{ccc}
Name & $N_\mathrm{gas}$ & Comments \\
\hline\hline
30k &     $3.0\times 10^4$ & Identical initial conditions files (ICfs).\\
30k(rd) & $3.0\times 10^4$ & Resampled ICfs (different Poisson noise).\\
50k &     $5.0\times 10^4$ & Identical initial conditions files.\\
75k &     $7.5\times 10^4$ & Identical initial conditions files.\\
100k &    $10.0\times 10^4$ & Identical initial conditions files.\\
\hline
\end{tabular}  
\end{table}
%
%
%
We use the models listed in table \ref{tabnpart} to investigate
the dependence of the models on the gas-particle number. These models
are all different realisations of the same set-up: that of the C05
model (see table \ref{tabm}). The 30k(rd), 50k and 75k sets consist of
5 simulations; the 30k set of 6 simulations and the 100k set of 3
simulations. The simulations of the 30k set all start from the same
initial conditions file, only the random number seed of the stochastic
star formation recipy (eq. (\ref{schmidt2})) varies between
simulations. The same goes for the 50k, 75k and 100k sets of
simulations. For the 30k(rd) set, the initial conditions file was
resampled for each simulation, i.e. they have variable Poisson noise
at constant gas-particle number. The number of dark matter particles
was kept constant at $30\,000$ particles for all simulations.

A first indication of the global behaviour of the models is shown in
Fig. \ref{npart_sfr} where we plotted the mean SFRs of the sets of
models listed in \ref{tabnpart}, with $1\sigma$ ``errorbars''
quantifying the variation between the simulations belonging to each
set. As a reference, the SFR of the C05 model is also
plotted. Obviously, the global SFRs are qualitatively and
quantitatively in good agreement and they do not change significantly
with gas-particle number. Of course, slight differences between the
different sets of models during the first star-formation peaks are
amplified towards the later peaks.
In figure \ref{npart_statistics} we plot different observables as a
function of stellar mass for the models listed in table \ref{tabnpart}
as well as the C05 model. We also plot a line connecting the C-series
models. Apparently, simulations with higher numbers of gas particles
tend to form slightly more stars, the difference between the 30k and
75k sets of simulations being of the order of 10\% in terms of stellar
mass and with comparable differences on the other observables. There
is however no straightforward converging behaviour as the 100k set
forms on average less stars than the 75k set. The least well
constrained quantity is the percentage of retained gas, with a 30\%
increase between the 30k and 75k models. Clearly, the scatter between
the individual models diminishes towards larger gas-particle numbers
but increasing the gas particle number beyond $75\,000$ seems to have
no further effect. We conclude that if one wants to compare an
individual model in detail to an observed galaxy, it seems advisable
to use at least $100\,000$ gas particles in order to minimize the
model-to-model scactter. For a comparison between observed and model
photometric and kinematic scaling relations, a gas-particle number of
$30\,000$ should suffice. The model-to-model scatter is sufficiently
small and, in any case, this scatter moves models along the scaling
relations so the latter will not be significantly affected. This
conclusion is a lot stricter than the one reached in \cite{lia2000},
where the requirement of comparable star formation histories resulted
in $N_\mathrm{gas}\ge10\,000$.
%
%

\begin{figure*}
\includegraphics[width=\hsize]{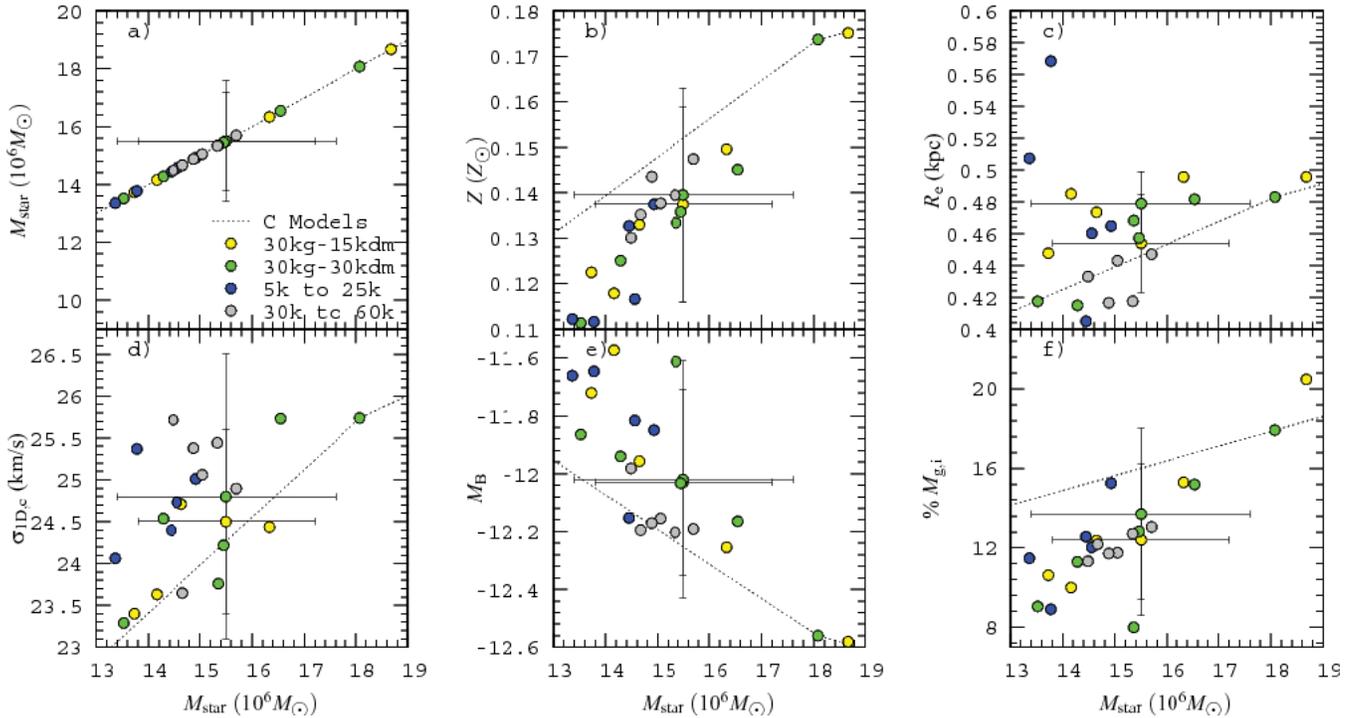}
\caption{Variation of model observables as a function of the dark
matter particle number. The dash-dotted line connects the points of
the C models. The symbol at the center of the error-bars indicates to
which models these bars belong (it is not a datapoint). The leftmost
two blue points in each figure are the models with $5\,000$ and
$10\,000$ DM particles (from left to right). For clarity: the larger
of the two sets of error-bars in each figure belong to the 30kg-30kdm
models. See table \ref{tabnpart_dm} for more information about the
models.}
\label{npart_statistics_dm}
\end{figure*}

To assess the number of dark matter particles to be used in the
simulations, we constructed a further series of models (see table
\ref{tabnpart_dm}). For the 15kdm we started from the same
initial conditions file and used 5 different random number seeds for
the star formation recipy. The same was done for the 30kdm
models. For the 5k-to-25k and 30k-to-60k models the number of dark
matter particles was varied whilst retaining the initial conditions
for the gas particles. Only 1 run per initial conditions file was done
for these models. Results are shown in
Fig. \ref{npart_statistics_dm}. Two important conclusions can be drawn
from this figure: (i) there is no significant decrease of the scatter
on the individual models as a result of an increase in dark matter
particles and (ii) DM particle numbers of $15\,000$ and up seem to be
sufficient to capture enough physics to closely follow the model
relations. Indeed, apart from the leftmost two blue points in each
plot (which correspond to $5\,000$ and $10\,000$ DM particles) all the
models follow the C model relations. Although we do not rule out a
decrease of the scatter on individual models as the number of dark
matter particles is further increased, the above two results allow us
to use $30\,000$ dark matter particles to investigate the relations
our model galaxies follow.

{ Moreover, we performed a series of simulations with C09 initial
  conditions but with the number of gas particles increased from
  $30\,000$ to $40\,000$, $50\,000$, and $60\,000$. Due to star
  formation, the 60k run reached a total of $200\,000$ particles after
  5~Gyr and became impracticably slow and we decided to stop
  it. Comparing these models, we found that no observable scales
  directly with the number of gas particles and that the
  model-to-model scatter on the observables is too small to affect our
  results and conclusions.  }

\begin{table}
\caption{Information concerning the simulations with varying dark
  matter particle numbers. ($N_\mathrm{gas}=30\,000$)}
\label{tabnpart_dm}
\centering
\begin{tabular}{cccc}
Name & $N_\mathrm{DM}$ & \# runs \\
\hline\hline
15kdm & $1.5\times 10^4$ & 5 \\
30kdm & $3.0\times 10^4$ & 5 \\
5k-to-25k & $(0.5,1,1.5,2,2.5)\times 10^4$ &  \\
30k-to-60k & $(3,3.5,4,4.5,5,6)\times 10^4$  \\
\hline
\end{tabular}  
\end{table}
In conclusion: for our production run models (C and D models) we use
$30\,000$ gas particles and $30\,000$ dark matter particles. These
numbers used in are an improvement over \citet{chiosi2002}, who use
$10\,000$ particles for dark and baryonic matter. They are less than
the $100\,000$ particles for DM and BM used by \citet{stinson2007},
but we have nevertheless demonstrated that this resolution is
sufficient to capture the essentials of the physics.
\subsection{Surface brightness Profile}
\label{subsec:sb}
\begin{figure}
\includegraphics[width=\hsize]{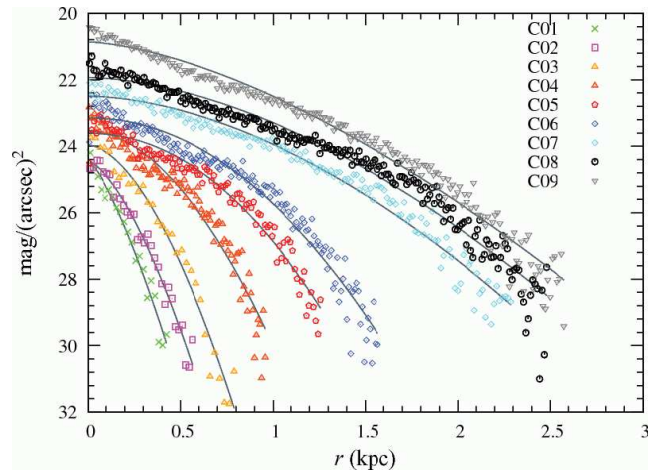}
\caption{Surface brightness (SB, in mag arcsec$^{-2}$) as a function
of radius (kpc) for simulations with varying initial mass. The full
lines are the best fitting S\'ersic profiles.}
\label{m_sersic}
\end{figure}

\begin{figure}
\includegraphics[width=\hsize]{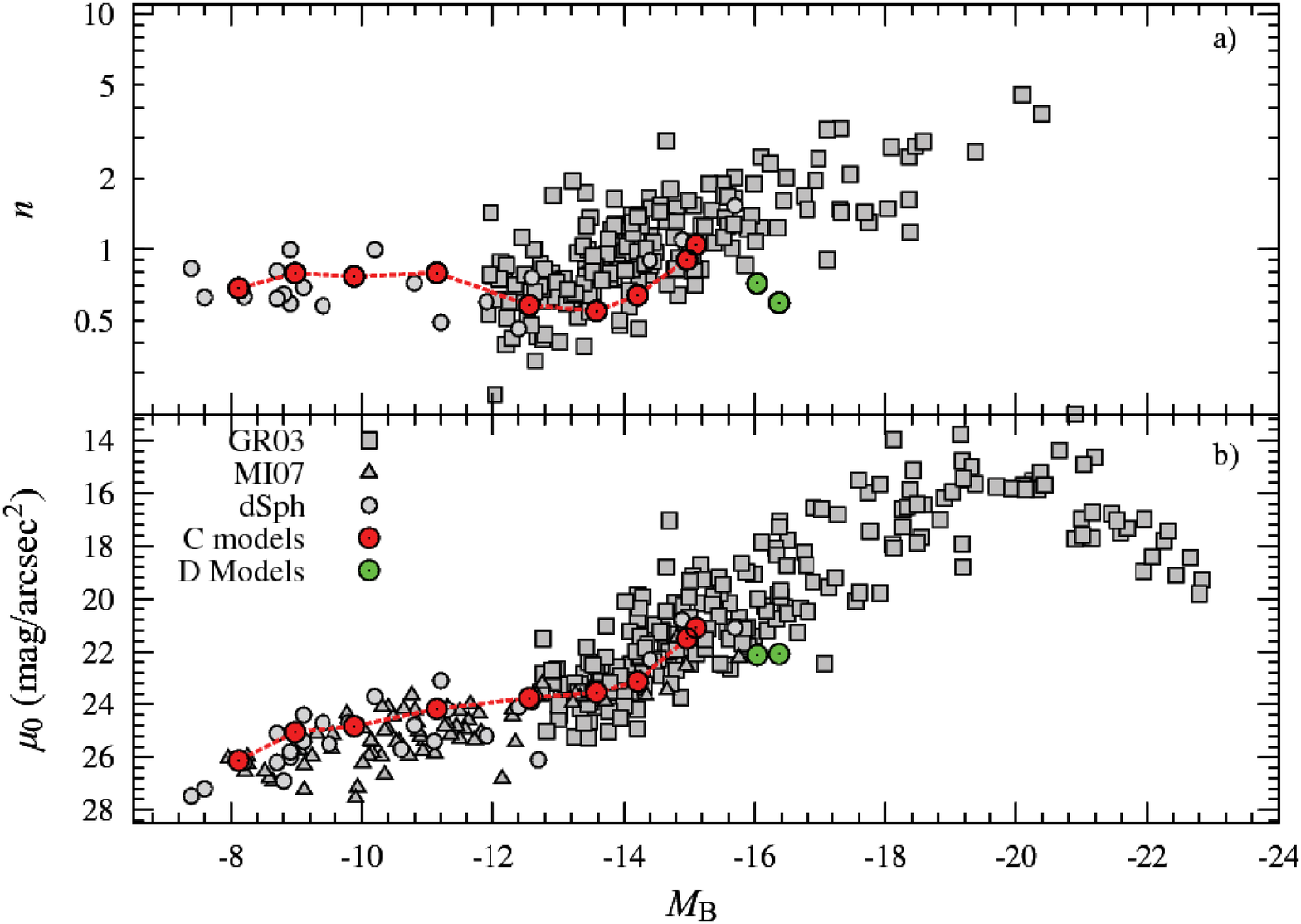}
\caption{\emph{Upper panel:} S\'ersic index $n$ as a function of blue
broad band magnitude $M_\mathrm{B}$. \emph{Lower panel:} central
surface brightness $\mu_\mathrm{0}$ (mag arcsec$^{-2}$) from the
S\'ersic fit to the models as a function of $M_\mathrm{B}$. Data taken
from \citet{graham2003}, \citet{derijcke2008} and (lower panel only)
\citet{mieske2007} (the canonical value of 0.7 was added to absolute
magnitudes and surface brightnesses in the $V$ band to mimic their
respective counterparts in the $B$ band).}
\label{m_ser_n_data}
\end{figure}

\begin{table*}
\begin{minipage}{155mm}
 \caption{Summary of various physical quantities for models C and D
 (at $t=10$ Gyr). (1) Model name (2) Initial gas mass ($10^6\
 M_{\sun}$) (3) Initial dark matter mass ($10^6\ M_{\sun}$) (4) $a$
 for the DM halo (kpc)(see eq. (\ref{kuzkutrho1})) (5) Absolute
 magnitude in the blue band (6) Absolute magnitude in the visual band
 (7) $B-R$ (8) Total final star mass ($10^6\ M_{\sun}$) (9) Percentage
 of initial gas mass remaining ($<$ 30 kpc) (10) Percentage of gas blown
 out ($>$ 30 kpc) (11) Half-light radius (kpc) (12) One dimensional
 central velocity dispersion (km s-1) (13) Metallicity
 ($Z_{\sun}=0.02$)}
 \label{tabm} \centering \begin{tabular}{ccccccccccccc} Run &
 $M_{\mathrm{g,i}}$ & $M_{\mathrm{DM,i}}$ & $a$ & $M_\mathrm{B}$ &
 $M_\mathrm{V}$ & $B-R$ & $M_\mathrm{star}$ & \% $M_\mathrm{g,i}$ & \%
 BO & $R_\mathrm{e}$ & $\sigma_\mathrm{1D,c}$ & $Z(Z_{\sun})$ \\
\hline\hline
       C01 & 44  & 206 & 0.439 &  -8.12 &  -8.72 & 1.067 & 0.44 & 21.0 & 78.0 & 0.16 &  9.6 &   0.048 \\
       C02 & 52  & 248 & 0.466 &  -8.97 &  -9.51 & 0.981 & 0.81 & 19.5 & 78.9 & 0.18 & 11.5 &   0.052 \\
       C03 & 70  & 330 & 0.513 &  -9.89 & -10.44 & 0.997 &  2.0 & 15.6 & 81.6 & 0.23 & 14.0 &   0.045 \\
       C04 & 105 & 495 & 0.587 & -11.14 & -11.68 & 0.973 &  6.2 & 9.1  & 85.0 & 0.31 & 18.9 &   0.073 \\
       C05 & 140 & 660 & 0.646 & -12.56 & -13.03 & 0.878 &   18 & 17.9 & 69.1 & 0.48 & 25.7 &   0.174 \\
       C06 & 175 & 825 & 0.696 & -13.58 & -14.02 & 0.831 &   38 & 32.7 & 45.7 & 0.67 & 31.5 &   0.230 \\
       C07 & 262 & 1238 & 0.797 & -14.20 & -14.83 & 1.105 & 122 & 36.6 & 16.5 & 0.76 & 39.3 &   0.379 \\
       C08 & 349 & 1651 & 0.877 & -14.97 & -15.56 & 1.043 & 228 & 22.7 & 11.5 & 0.62 & 43.0 &   0.535 \\
       C09 & 524 & 2476 & 1.004 & -15.11 & -15.77 & 1.162 & 393 & 17.4 & 6.80 & 0.57 & 46.7 &   0.620 \\
&&&&&&&&&&\\
       D01 & 873 & 4127 & 4.000 & -16.04 & -16.61 & 1.023 & 579 & 29.0 &  4.2 & 1.13 & 35.2 &   0.470 \\
       D02 & 873 & 4127 & 6.000 & -16.37 & -16.82 & 0.843 & 488 & 39.4 &  4.4 & 1.26 & 30.9 &   0.373 \\
 \hline
 \end{tabular}
\end{minipage}
\end{table*}
The surface brightness (SB) profile of elliptical galaxies is
described well by the de Vaucouleurs $R^{1/4}$ law, whereas dwarf
elliptical galaxies are fitted better by an exponential law \citep[see
e.g.][and references therein]{jerjen1997,graham2003}. Both profiles
are encompassed by S\'ersics $R^{1/n}$ law:
\begin{equation}
I(R) = I_0 \exp{\left\{-\left(\frac{R}{R_0}\right)^{\left(1/n\right)}\right\}}.
\label{sersic1}
\end{equation}
In Fig. \ref{m_sersic} we present a plot of the surface
brightness profiles of the C model galaxies as well as their
respective best fitting S\'ersic profiles. In Fig. \ref{m_ser_n_data}
we compare the S\'ersic parameters $n$ and $\mu_\mathrm{0}$ (or
$I_\mathrm{0}$) for the models with data taken from
\citet{graham2003}, \citet{derijcke2008} and \citet{mieske2007}. As
seen in Fig. \ref{m_sersic}, the correspondence with a S\'ersic
profile is excellent. The more massive models ($M_\mathrm{B}<-13$ mag)
also have SB profiles that closely follow the S\'ersic law. Comparison
with data in Fig. \ref{m_ser_n_data} for the C models shows that $n$
as well as $\mu_\mathrm{0}$ are in good correspondence with
observations. A variant of the third S\'ersic parameter
$R_\mathrm{0}$, the half-light radius $R_\mathrm{e}$, is discussed in
section \ref{subsec:halfl}. The D models perform worse than the C
models~: whilst $\mu_\mathrm{0}$ is only somewhat too low, the values
for $n$ are too low. This indicates that the SF should be more
centrally concentrated. This is easily understood when considering
Table \ref{tabm}: by construction the D models have a large radius,
naturally decreasing the resulting $\mu_\mathrm{0}$ and $n$.
%
\subsection{Star Formation Rate}
The SFR of galaxies is an important tool for the study of their
evolution. Observations indicate that the Star Formation Histories of
dwarf galaxies are varied and have a burst-like nature
\citep[e.g.][and references therein]{smeckerhane1996,mateo1998}. This
phenomenon was successfully reproduced by N--body simulations
\citep{chiosi2002,stinson2007}. \citet{dellenbusch2007} find ongoing
star formation in their sample of observed dwarf galaxies. They also
note and an absence of any indication of a recent major
interaction. These observations can be readily explained by invoking
the burst-like nature of star formation in dwarf galaxies. 

\begin{figure}
  \includegraphics[width=\hsize]{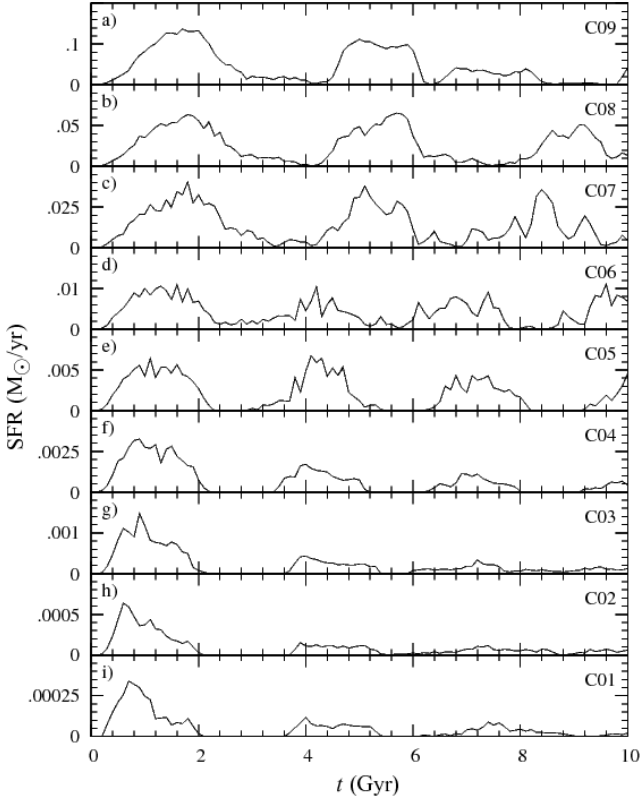}
\caption{SFR (M$_{\sun}$ yr-1) as a function of time (Gyr) for the C models.}
\label{m_sfr}
\end{figure}

In Fig. \ref{m_sfr} we plot the SFRs of the C models.
%
%
We immediately see that the SFR is of a burst-like nature. As there is
10 to 30 per cent of the initial gas mass left at the end of the
simulations (Table \ref{tabm}) there is still fuel for further star
formation. Most dwarf elliptical (dE) galaxies are found to be almost
devoid of gas \citep{young1997,mateo1998, conselice2003, derijcke2003,
  bouchard2005, buyle2005}. There is however a very strong
morphological segregation of dwarf elliptical galaxies
\citep{binggeli1990}. These authors found that almost all dEs are
found in clusters or in dense groups, or as companions (satellites) of
a massive galaxy. This implies that dwarf elliptical galaxies, found
in dense environments, are subject to significant ram-pressure
stripping which depletes their gas reservoirs
\citep{mori2000,marcolini2003}. This physical process is not included
in our simulations so the modelled galaxies retain part of their gas
reservoir.

\begin{figure}
\includegraphics[width=\hsize]{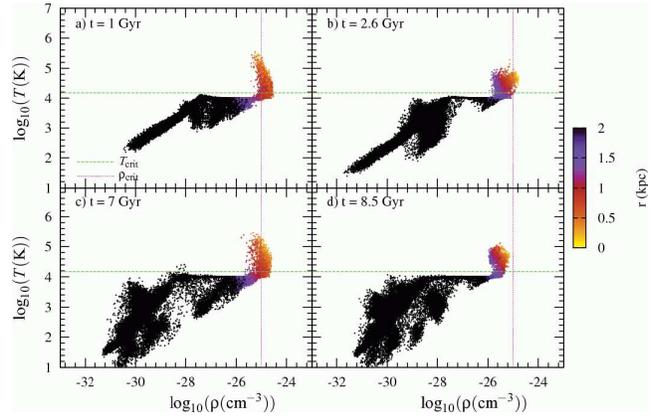}
\caption{{ The logarithm of the temperature, $T$, versus the
    density, $\rho$, for the gas particles, at 4 different times (C05
    model). Particles are colour-coded according to their distance
    from the galaxy centre $r$.}}
\label{temperature_density}
\end{figure}

{ In Fig. \ref{temperature_density} we plot the temperature
  distribution of the gas particles as a function of their density,
  for the C05 model, at 4 different times: 1, 2.6 , 7 and 8.5 Gyr. In
  each figure the lower right quadrant contains the gas particles that
  satisfy the density and temperature criterion for star formation
  (eqs. (\ref{sfcrit2}) and (\ref{sfcrit3})). From Fig. \ref{m_sfr} we
  can see that at 1 and 7 Gyr the galaxy is undergoing a
  star-burst. At 2.6 and 8.5 Gyr there is no star-formation. From
  Fig. \ref{temperature_density} it is clear that the quiescent
  intervals between star-bursts are caused by a decrease in gas
  density. This in turn is a result of the stellar feedback: the
  entire central population of gas particles is pushed to lower
  densities, inhibiting further star formation. On the lower left part
  of each figure we can see gas particles being blown out of the
  galaxy, resulting in a declining density and temperature. On panels
  b), c) and d) we can clearly see the outward propagation of the gas
  particles expelled by the different starbursts towards the lower
  left corner of the figure.}

\begin{figure}
\includegraphics[width=\hsize]{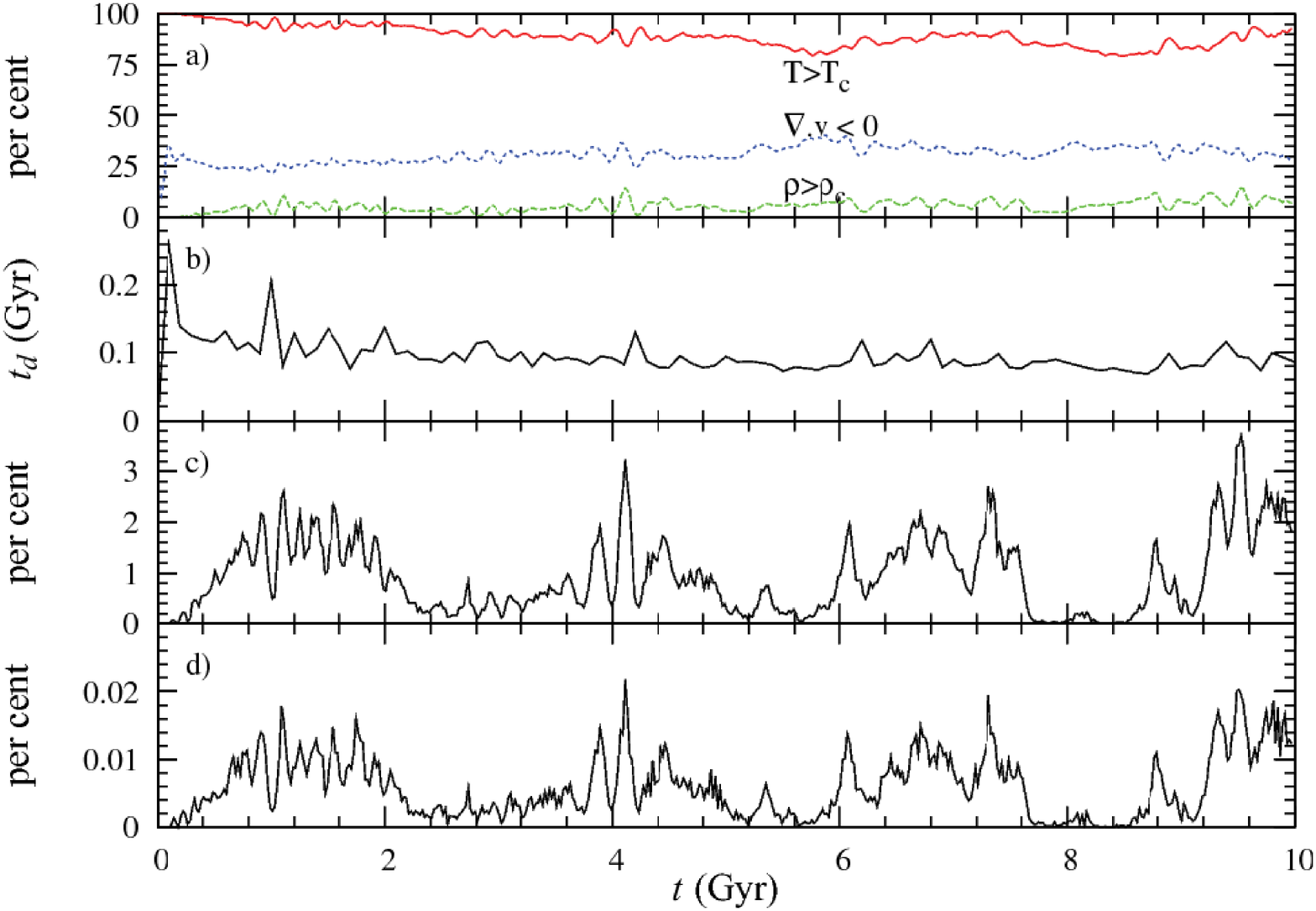}
\caption{Data taken from the C06 model. {a)} Percentage of gas
particles satisfying a SF criterion. {b):} Average dynamical time
(Gyr) for gas particles within 0.5 kpc of the galaxy centre. {c)}
Percentage of gas particles satisfying the 3 SFC. {d)} Percentage of
gas particles forming stars.}
\label{C06_sfc}
\end{figure}

The cause of the burst-like behaviour can be seen on Fig.
\ref{C06_sfc}, where the effects of the various SFC are plotted for
model C06. From the upper panel we learn that the temperature
criterion (eq. \ref{sfcrit1}) has little influence on SF, prohibiting
a small fraction of the gas particles from forming stars. The
divergence criterion (eq. \ref{sfcrit2}) is more restrictive, allowing
on average 30 per cent of the gas particles to form stars and restricting the
star formation to the more central regions of the galaxy. The real
control over star formation resides however in the density criterion
(eq. \ref{sfcrit3}). Not only is this criterion very restrictive
(which of course depends on the value of $\rho_c$), typically allowing
5--10 per cent of the gas particles to form stars, it is also responsible
for the burst-like behaviour the star formation. Star formation is in a
sense a self-regulating process: high density gives high SF, resulting
in a lot of SNe which will heat up and disperse the gas, leading to
the end of the SF episode. As the temperature criterion has only
little effect it could be dropped altogether (no unrealistic star
formation should take place in hot regions because hot clouds with
high density will cool fast anyway, hot clouds with low density are
not allowed to form stars). \citet{kravtsov2003,li2005a,li2006} use
only the density criterion in their N--body/SPH simulations, while using
enough particles to satisfy the Jeans resolution criterion
\citep{bate1997} and thus resolving Jeans instability in molecular
clouds. They find that the Schmidt law (eq. \ref{schmidt1}) is a
result of gravitational instability and thus of the overall density
distribution, which validates our SF because we also find it to be
--mainly-- determined by the density distribution.
When comparing panels a) and b) of Fig. \ref{C06_sfc} we see the
behaviour expected from $t_d$ (equation (\ref{dyntime1})): peaks when
there are few particles with high density and vice verse, with an
approximately constant value otherwise ($\approx 0.12$ Gyr). When
going from panel 3) to panel 4), equation (\ref{schmidt2}) is
applied. Because of the moderate variations of $t_d$ there is little
qualitative difference between the two panels, although the
quantitative difference is large (factor 100). The use of equation
(\ref{schmidt2}) is thus twofold: on the one hand it serves as a means
to drastically reduce the number of star forming gas particles, on the
other hand it introduces an independence of the number of time-steps
into the star formation prescription.

\begin{figure}
\includegraphics[width=\hsize]{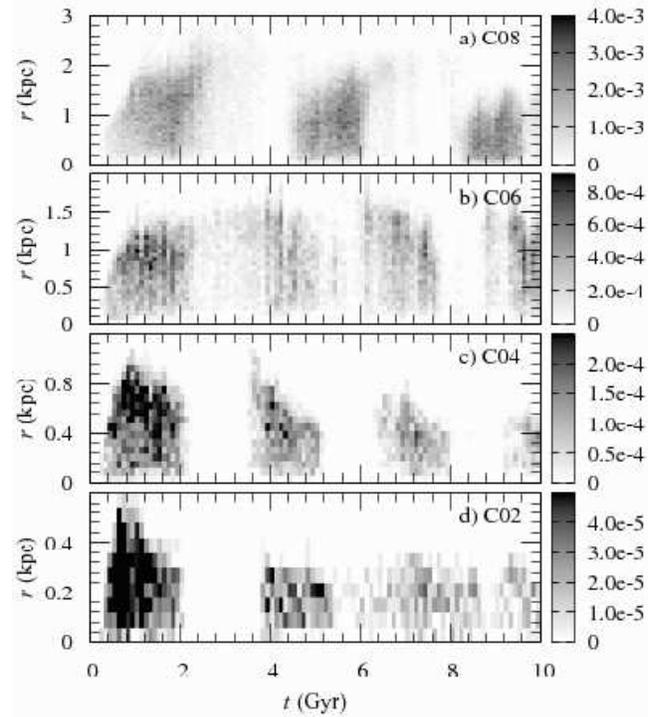}
\caption{SFR ($M_{\sun}$/Gyr) as a function of time and distance from
the galaxy centre, for four different models. The SFR is integrated
over spherical shells with width $r\rs{max}/50$, with $r\rs{max}$ from
the top to the bottom panel: 3,2,1.2,0.6 kpc respectively.}
\label{sfrr3dall}
\end{figure}

\begin{figure}
\includegraphics[width=\hsize]{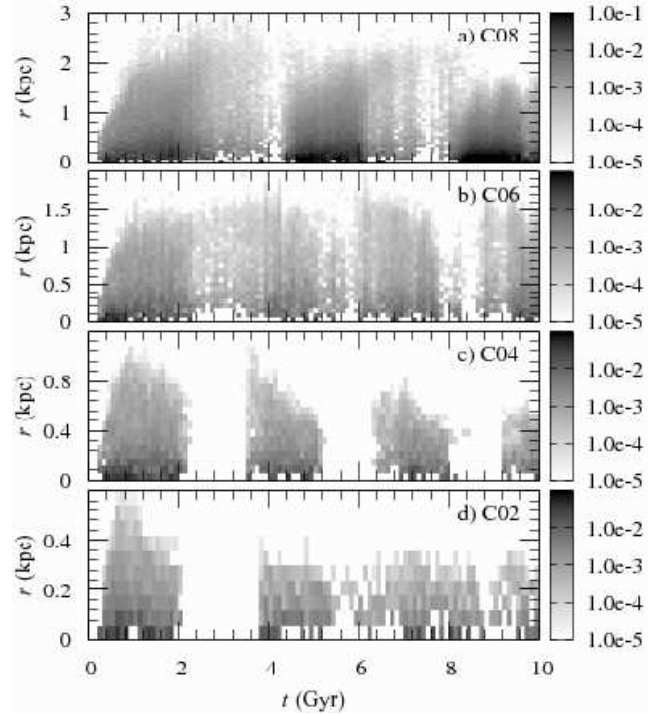}
\caption{SF density ($M_{\sun}$/Gyr/kpc$^3$) as a function of time and
distance from the galaxy centre, for four different models. The
spherical bins are equal to those of Fig. \ref{sfrr3dall}.}
\label{sfrr3dalldens}
\end{figure}

In Fig. \ref{sfrr3dall} we show a measure of the distribution of SF in
space for our models. The apparent rise of the SFR towards larger
radii (within a starburst) is caused by the rise in volume of the
spherical bins. The volume-averaged SFR, plotted in
Fig. \ref{sfrr3dalldens}, shows that there is actually more SF in the
centre of the Model galaxies. We can also see that in our more massive
models, during the first starburst, SF takes an increasing amount of
time to reach the outskirts of the galaxy. On panels a), c) and d) of
Fig. \ref{sfrr3dall} we see that star formation becomes more centrally
concentrated as time passes: each subsequent star-burst is confined to
a smaller radius. For model C02 (panel d)) this effect is even
stronger: the first star-burst contains two phases, with SF in the
first phase extending to about $\approx$ 0.5 kpc, in the second phase
to $\approx$ 0.3 kpc. Our models thus naturally reproduce the
observations of \citet{tolstoy2004}. These authors found two distinct
ancient components in the Sculptor dSPH, the metal-poor component
being more extended in space. They considered the interplay of gas
infall due to gravitation and supernova feedback to be a possible
explanation. \citet{battaglia2006} found a similar effect in the
Fornax dSph: they found 3 distinct stellar populations (ages $>10$
Gyr, 2--8 Gyr and $< 100$ Myr), where the younger, more metal-rich
populations are found to occupy a smaller region of space. We can
immediately qualitatively compare these findings with our C04 Model,
which also gives rise to 3 different populations, with each subsequent
population being more metal-rich and confined to a smaller region in
space. \citet{mashchenko2007} offer a different explanation: the DM,
gravitationally heated because of gas motion, in turn heats the
stellar population. As the older (metal poor) population has somewhat
more time to be heated than the metal rich population, it will occupy
a larger region in space. These authors, however, do not provide
quantitative predictions based on this scenario. Note that the
apparent absence of SF in the inner radial bin, visible in Fig
\ref{sfrr3dalldens}, is an artifact caused by small number statistics
as the size of the volume over which we calculate the star-formation
density becomes small compared with the inter-particle distances.

\begin{figure}
\includegraphics[width=\hsize]{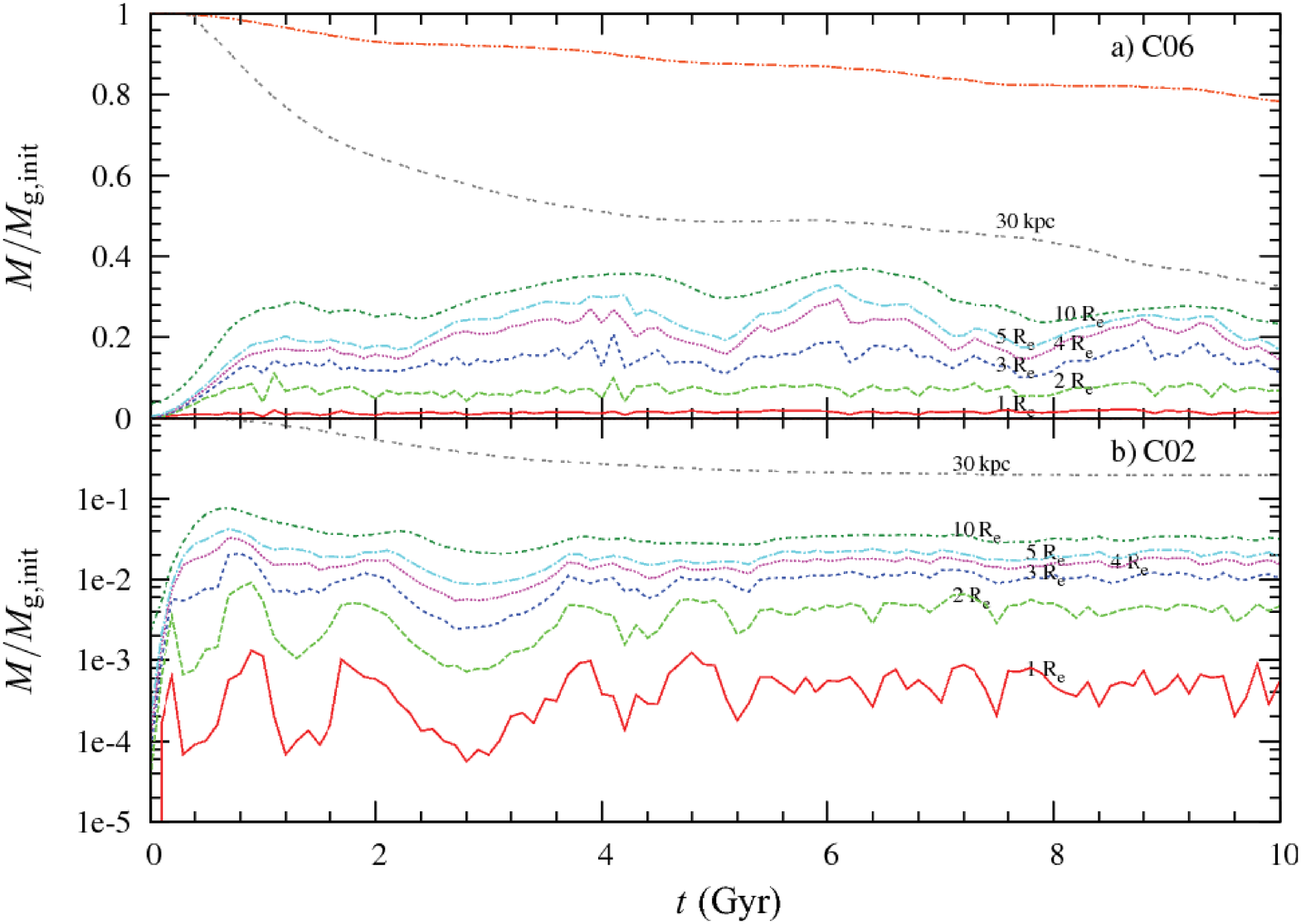}
\caption{\label{cummassfig}Evolution of the gas mass as a function of
  time (Gyr), for Models C06 (upper panel) and C02 (lower panel). The
  different lines show the fraction (mass divided by the initial gas
  mass) of the gas mass that is contained in a sphere with a given
  radius ($1R\rs{e}, 2R\rs{e},3R\rs{e}, 4R\rs{e}, 5R\rs{e},
  10R\rs{e},30\ \mathrm{kpc}$). The top line shows the fractional gas
  mass remaining (i.e. not converted into stars.) \re is the
  half-light radius at the end of the simulation (see Table
  \ref{tabm}). The top line for panel b) is not visible because it is
  $\approx 1$.}
\end{figure}

From Table \ref{tabm} we can conclude that blow-out of gas in the
models is not very efficient. Although the least massive models easily
lose up to 80 per cent of their gas, this declines to around 5 per
cent for the most massive C models and for the D models. This
difference in dynamical evolution is shown in Fig. \ref{cummassfig},
showing the evolution of the gaseous component of 2 models, as a
function of time. While the C06 model is able to retain most of the
gas and steadily converts it into stars, the C02 model is clearly not
massive enough to be able to capture a large amount of
gas. \citet{maclow1999} find that blow-away occurs for
$M_\mathrm{gas}$ up to $10^5-10^6\ M_{\sun}$, and that no blow-out of
gas occurs for gas masses $\approx 10^8-10^{10}\ M_{\sun}$ (depending
on the energy released by SNe). These observations are broadly
consistent with our results: gas loss is very small for initial gas
masses $\geq 4\times 10^{8}\ M_{\sun}$. Our upper limit for blow-away
is $\approx 4\times 10^7 M_{\sun}$ (C01 model). \citet{ferrara2000}
find a comparable limit for blow-out as \citet{maclow1999}:
$M_\mathrm{gas}\approx 10^8-10^9\ M_{\sun}$, also in good agreement
with our results.
%
%
\subsection{Half-light radius}
\label{subsec:halfl}
\begin{figure}
\includegraphics[width=\hsize]{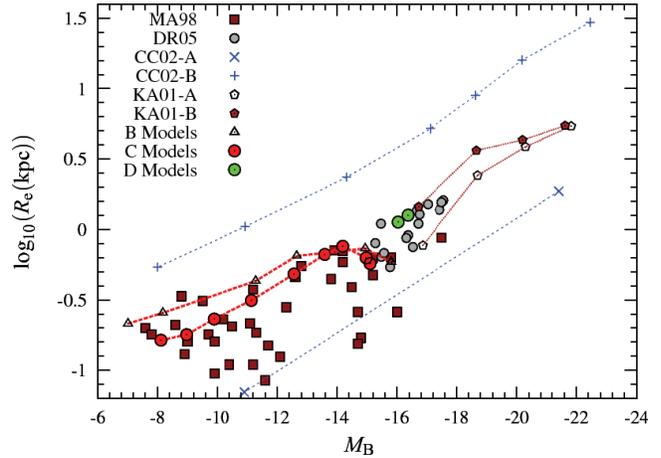}
\caption{Logarithm of the half-light radius \re as a function of the
absolute magnitude in the B-band $M_\mathrm{B}$ for several data-sets
(observations: MA98 \citep{mateo1998} (filled boxes: data that only
appears in this plot), DR05 \citep{derijcke2005}, theory: CC02
\citep{chiosi2002}, KA01 \citep{kawata2001}) as well as the models.}
\label{MB_re}
\end{figure}

The half-light radius, the radius of a galaxy containing half its
total luminosity, is an important diagnostic parameter in simulations
of galaxy formation. Large deviations from values dictated by
observations hint at an erroneous spatial distribution of the star
formation of model galaxies. Instead of using the three dimensional
half-mass radius for $R\rs{e}$, we chose to calculate the value
accessible to observers: the two dimensional half-luminosity
radius. We first projected the light distribution of the models onto
the $xy$ plane and derived the cumulative luminosity function $L(R)$
as a function of radius. The value of $R$ where $L(R)$ reaches half
its maximum is then the half-light radius. Results for the models are
shown in Fig. \ref{MB_re}, where the logarithm of $R_\mathrm{e}$ is
shown as a function of $M_\mathrm{B}$. Throughout the simulations we
found that the half-light radius of a model galaxy is very sensitive
to the setup (e.g. initial conditions) of the simulations. Restraining
it to a reasonable value is a far from trivial task, as can also be
seen by the large variations in the values of the CC02 models (Fig.
\ref{MB_re}). Comparison of the data-points for the B and C models
show that increasing the size of the DM halo results in an increase of
the galaxy half-light radius. When the models are to be compared to
the CC02 models it should be to the CC02-B models, because those have
initial densities closely resembling ours (the over-density of matter
in the CC02-B models is set to 1:
$\rho_\mathrm{gas}=M_\mathrm{gas}/M_\mathrm{DM}\times\rho_\mathrm{u}\approx0.11\,\rho_\mathrm{u}$,
whereas the CC02-A models have
$\rho_\mathrm{gas}\approx22\,\rho_\mathrm{u}$. We have
$\rho_\mathrm{gas}=M_\mathrm{gas}/M_\mathrm{DM}\times5.55\times0.3\,\rho_\mathrm{u}\approx0.35\,\rho_\mathrm{u}$
(see eq. (\ref{densbar1}))). Although somewhat too large, the model
galaxies (B and C) perform reasonably well in the range
$M_\mathrm{B}\in[-6,-15]$ mag. When we look at the larger models
($M_\mathrm{B}<-15$ mag) we see that $R_\mathrm{e}$ decreases with
increasing mass, signalling a much too central concentration of star
formation compared with observations. A plausible explanation for this
failure is that the breakdown of the models signals where the effects
of hierarchical merging can no longer be neglected in the formation of
galaxies, and thus that the simple initial conditions are no longer
accurate in this region. When looking at the D models, where we chose
$a$ for the DM haloes to be an arbitrary larger value (Table
\ref{tabm}), we see that it is possible to produce massive dwarf
galaxies that do comply with observations, using a more extended halo
than dictated by equation (\ref{kka1}). Here the correspondence with
observations is excellent, showing not only that we can indeed regard
the extent of the halo (within our set framework of star formation,
feedback, ...) as the decisive factor concerning \re but also that we
could fine-tune it, without changing any other parameters, to get
excellent agreement between theory and observations over the entire
dwarf magnitude range $M_\mathrm{B}\in[-8,-18]$ mag for $R\rs{e}$. As
the discussion in section \ref{subsec:sb} shows, however, these models
deviate from the expected S\'ersic profile. This deviation could be
alleviated by changing (reducing) $a$. We feel however that fiddling
with parameters from model to model to be able to cover the entire
magnitude range of the dwarf galaxies with excellent precision would
not give much insight in the physics involved. A preferred strategy
would be to study the evolution of dwarf galaxies in a fully
cosmological setting, or to derive information about the initial
conditions (extent and shape of the DM haloes, the ratio of gas mass
to DM mass, ...) directly from numerical simulations. This is beyond
the scope of this paper. We limit ourselves in this paper to the mass
range where our chosen set of parameters gives good results. We note
that the SAMs of \citet{naga2005} seem to be suffering from the same
behaviour of $R\rs{e}$, it being rather large for small
galaxies. There it is argued that this is a result of either the
limited resolution of their N-body simulations, used to construct a
merger tree, or that the SF timescale at high redshift is too long.
\subsection{Colour}
\begin{figure}
\includegraphics[width=\hsize]{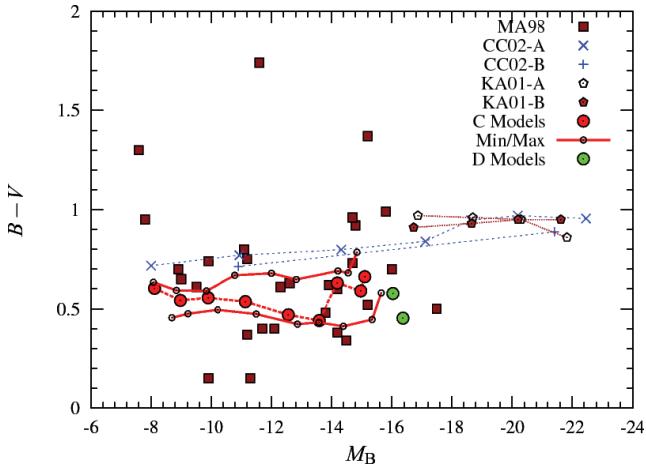}
\caption{$B-V$ colour of observed galaxies as well as the model galaxies as a function of blue magnitude $M_\mathrm{B}$.}
\label{mbmv-mb-plm}
\end{figure}

When comparing the $B-V$ colour of the models with observations (Fig.
\ref{mbmv-mb-plm}) we find good agreement. Besides the value of $B-V$
at $t=10$ Gyr we also plotted the minimum and maximum value in the
interval 7.4--10 Gyr for each simulation, as a function of
$M_\mathrm{B}$ at those times (open tilted squares). This gives an
idea of the variation of $B-V$ during the evolution of the model
galaxies. This variation is significant, which can be readily
understood when looking at e.g. Fig. \ref{m_sfr}, as star-bursts
tend to lower $B-V$ because of the formation of bright blue stars
\citep[see the discussion in][]{chiosi2002}. In this view the higher
values for the CC02 models can be explained by looking at their fig.
2: all models have 1 predominant SF peak, no SF takes place after 6
Gyr of evolution, thus excluding the presence of high-luminosity young
blue stars. Higher values of $B-V$ could be obtained for the models by
e.g. removing the gas content of a model after 1 or 2 SF peaks (as
would be expected to happen in a cluster environment). When taking the
state of Model C04 at 4 Gyr, which is between the first and second SF
peak, and calculating luminosities as if the galaxy has an age of 10
Gyr, thus mimicking quiet evolution after the first SF peak, we get a
$B-V$ value of 0.71, close to the CC02 values.
\subsection{Central velocity dispersion}
\label{subsec:vd}
\begin{figure}
\includegraphics[width=\hsize]{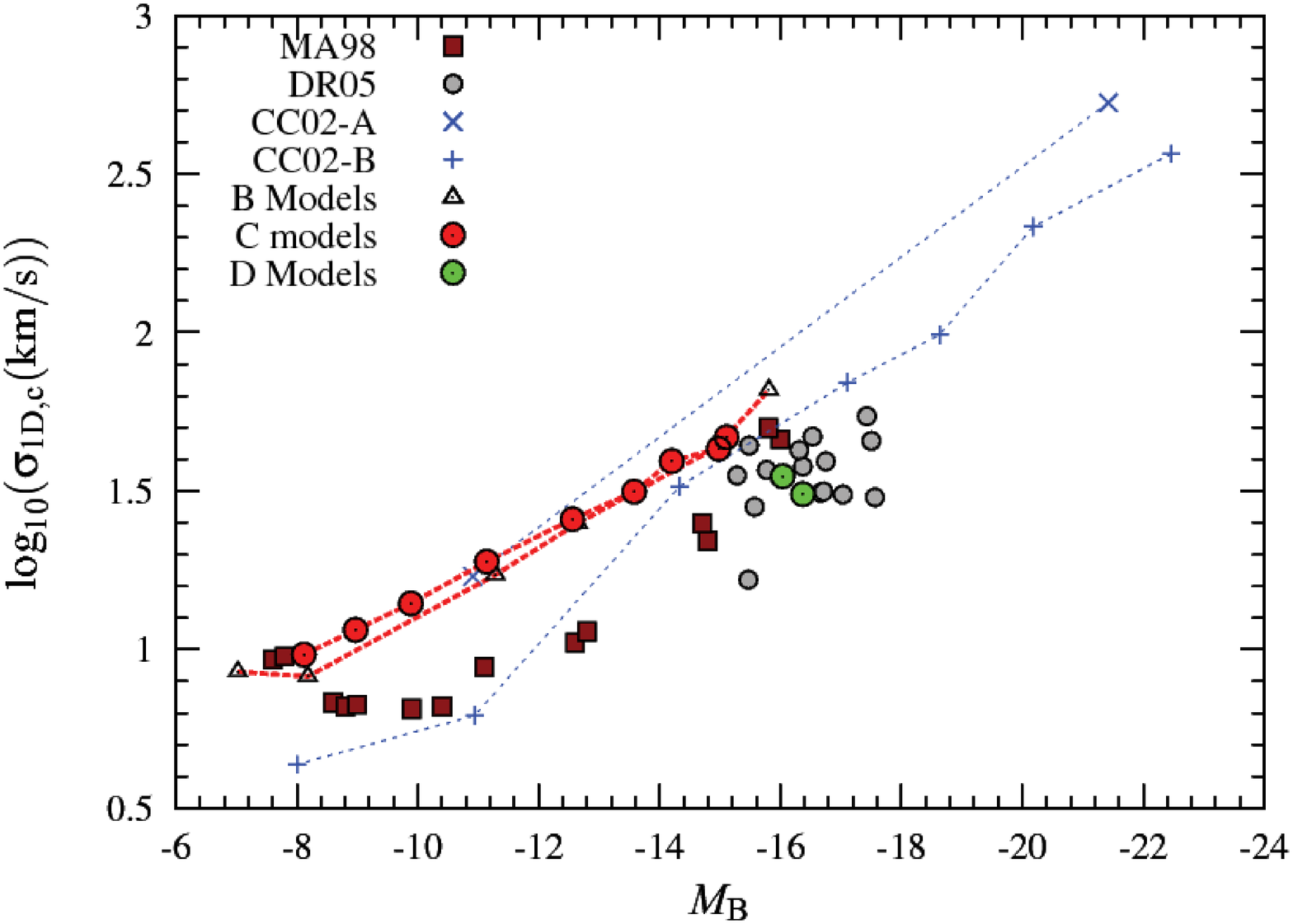}
\caption{Logarithm of the 1-dimensional central velocity dispersion
$\sigma$ (km s-1) as a function of the absolute magnitude in the B-band
$M_\mathrm{B}$ for several data-sets, as well as the models.}
\label{MB_disp}
\end{figure}

We used the one dimensional central velocity dispersion, which is the
three dimensional velocity dispersion ($\sigma$) divided by
$\sqrt{3}$. Although, from an observational point of view, there are
better measures for $\sigma$ (e.g. light weighted average $\sigma$)
because of possible nucleation of dwarf galaxies,
$\sigma_\mathrm{1D,c}$ is a good enough measure for our comparison. We
measured the central $\sigma$ by fitting an exponential function to
the dispersion profile, of which we retain the central value. Results
are shown in Fig. \ref{MB_disp}. Except for the least and the most
massive C models, $\sigma$ is too high. The correspondence for the D
models is however excellent. Overall we find reasonable convergence
between models and observations within a factor of 2 (0.3 dex). We further
note that a comparison of Figs. \ref{MB_re} and \ref{MB_disp} learns
that raising $a$ (going from the C to the B models) results in an
increase of the half-light radius as well as a decrease of central
$\sigma$. Because of this we cannot increase $a$ for the light models
to get a better fit as this would cause $R_\mathrm{e}$ to
increase. \emph{The combination of $R_\mathrm{e}$ and the central
$\sigma$ thus provide an excellent tool to evaluate the soundness of
halo properties within a certain framework of galaxy formation.}
\subsection{Metallicity}
\begin{figure}
\includegraphics[width=\hsize]{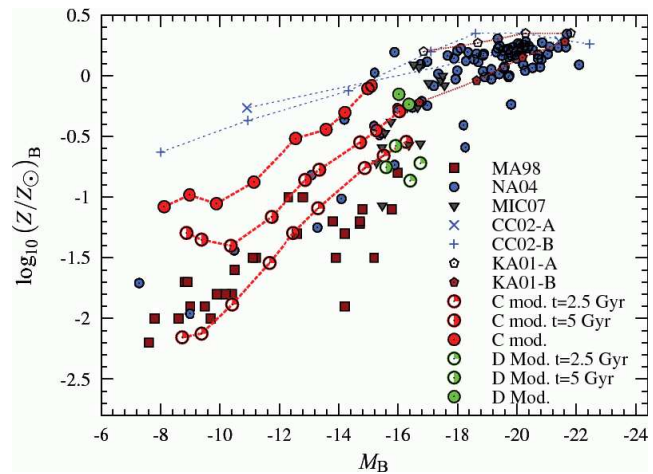}
\caption{Logarithm of the metallicity of model galaxies (plotted at
$t=2.5,5,10$ Gyr) as well as data-points (NA04 \citep{naga2004}, MIC07
\citep{michielsen2007}) as a function of blue magnitude. For the CC02
models the maximum metallicity is plotted.}
\label{Z_MB}
\end{figure}

In Fig. \ref{Z_MB} we plot metallicity
($\log_{10}(Z/Z_{\sun})\rs{B}$, with $Z_{\sun}=0.02$), for the C and D
models as well as 3 sets of data-points. $Z$ for our models is
$Z\rs{B}$, the light-weighted blue band metallicity. New and improved
information concerning the age and metallicity of dwarf galaxies has
become available during the recent years
\citep{geha2003,vanzee2004,michielsen2007,penny2007}. Dwarf galaxies
appear to be young, with $[Z/\mathrm{H}]$ approximately between -0.5
and 0, and a solar $[\alpha/\mathrm{Fe}]$. Three sets of points are
plotted for each model, respectively at $t=2.5,5$ and 10 Gyr. For the
CC02 models the maximum metallicity is plotted. The overall trend is
that metallicity increases with galaxy (star) mass \citep[see
e.g.][]{tremonti2004}. We see from Fig. \ref{Z_MB} that the C models
reproduce the observed trend well. The values agree reasonably with
the data-points, being on average 0.2 dex too high. For the D models
the correspondence with the observations is good. At 2.5 and 5 Gyr all
the models correspond excellent with the data-points.
%
\begin{figure}
\includegraphics[width=\hsize]{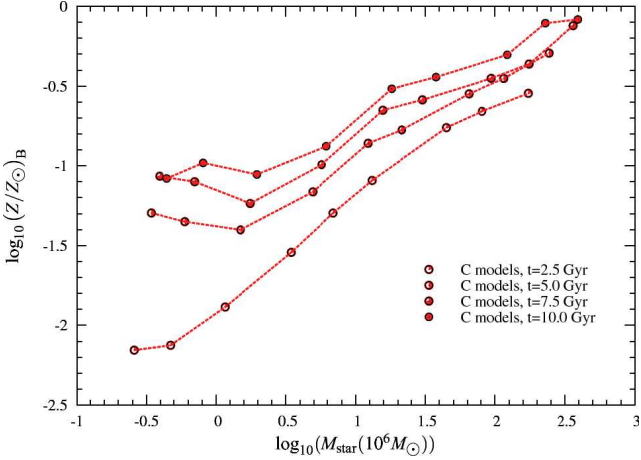}
\caption{Logarithm of the metallicity $Z\rs{B}$ of model galaxies as a
function of the logarithm of star mass at 4 different times (from
bottom to top: 2.5,5,7.5,10 Gyr) for the C models.}
\label{mstar-Z}
\end{figure}

In Fig. \ref{mstar-Z} we plot $Z$ as a function of stellar mass at 4
different times for the C models. The initial decline of $Z$ with
stellar mass is a consequence of the rise of the number of SNe, which
is not compensated for by the deepening of the galaxy potential
well. This is confirmed by Table \ref{tabm}, where the remaining gas
mass is seen to drop when going from model C01 to C03.
From the C03 model upwards, the metallicity rises with increasing
galaxy mass, indicating that the potential well is steep enough to
compensate for the increase in energy input into the ISM. No clear
trend concerning the remaining gas of these models can be seen in
Table \ref{tabm}. 

This effect is not visible in the observational data. It is however
unlikely that the decrease of metallicity with stellar mass would be
observable in dwarf spheroidal galaxies as it depends strongly on the
star formation in the range [3-10] Gyr (fig. \ref{m_sfr}), and we
expect that external processes (e.g. stripping) will greatly affect
the evolution of these light systems. Furthermore this effect is model
dependent and might be attenuated or even erased when modelling
galaxies using a different DM halo.
%
%
\subsection{The fundamental plane}
\begin{figure}
\includegraphics[width=\hsize]{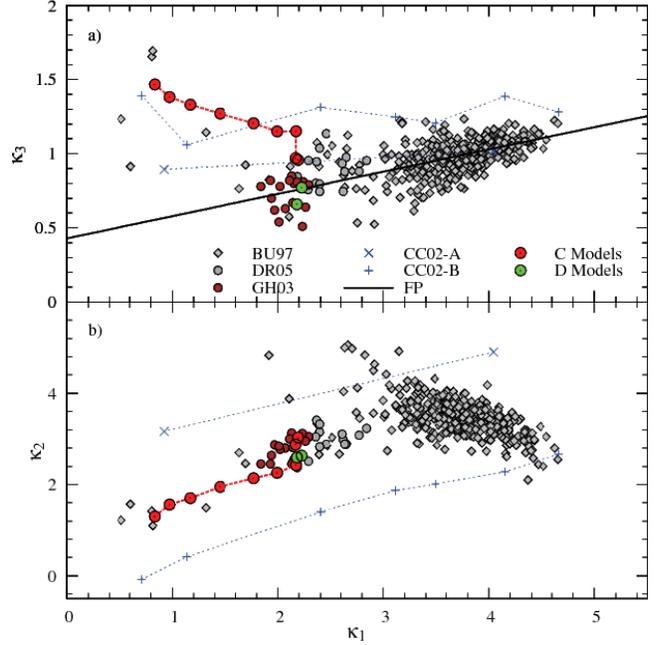}
\caption{The fundamental plane in $\kappa$-space
($\kappa_1,\kappa_2,\kappa_3$.). a) edge-on. The solid line is the
equation of the FP derived from Virgo galaxies: $\kappa_3 =
0.15\,\kappa_1 + 0.43$ \citep[][corrected to $H_0=70$ km s$^{-1}$
Mpc$^{-1}$]{burstein1997}. b) face-on. Data represented by open
squares are taken from \citet{burstein1997}, data represented by
asterisks are taken from \citet{geha2003}.}
\label{FPkappa}
\end{figure}

\begin{figure}
\includegraphics[width=\hsize]{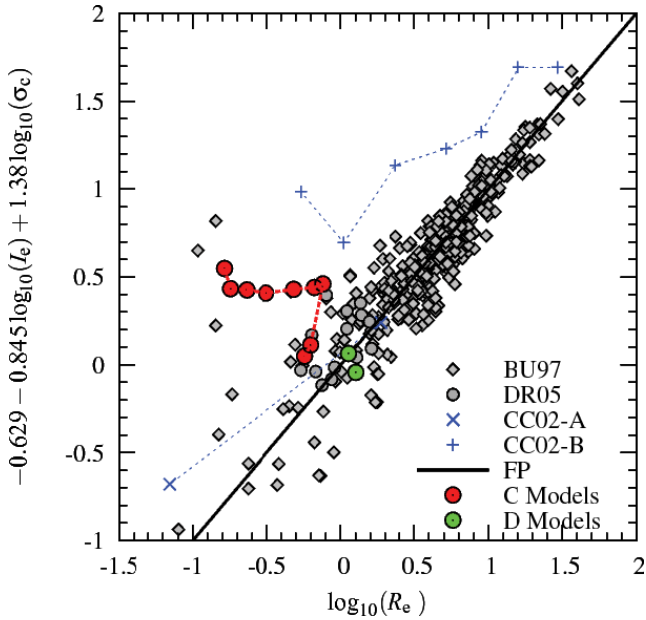}
\caption{The fundamental plane in physical coordinates:
$\log_{10}(\re)=-0.629-0.845\log_{10}(I\rs{e})+1.38\log_{10}(\sigma\rs{c})$. The
full black line shows a side projection of the FP.}
\label{FP2_1}
\end{figure}

\begin{figure}
\includegraphics[width=\hsize]{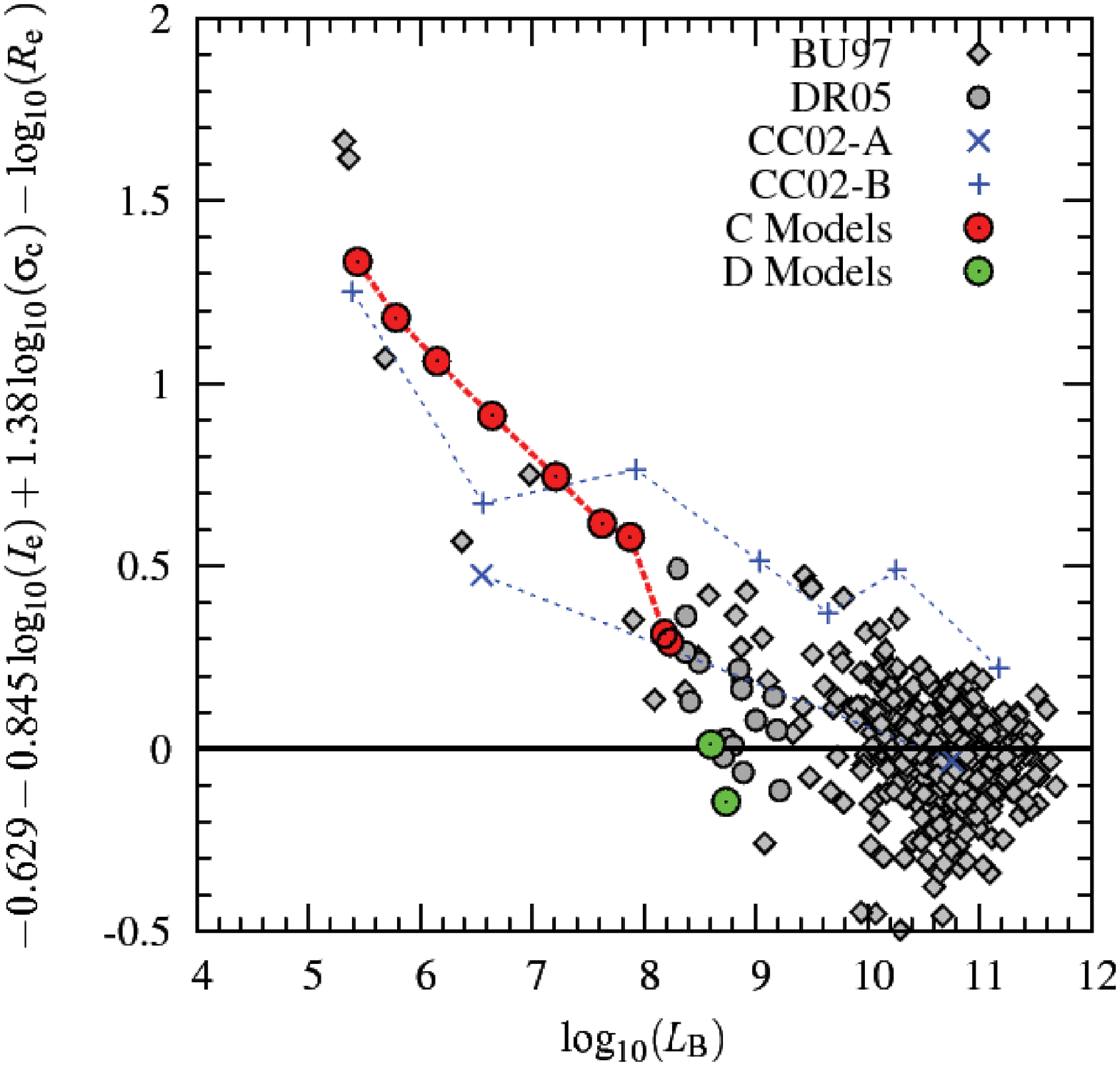}
\caption{Plot of the deviation from the fundamental plane. The black
line represents zero deviation from the FP.}
\label{FP2_2}
\end{figure}

The fundamental plane (FP) in $\kappa$-space \citep{bender1992}, as defined by the Virgo galaxies, is given by \citep{burstein1997}:
\begin{equation}
\kappa_3 = 0.15\,\kappa_1+0.43.
\end{equation}
where
\begin{eqnarray}
\kappa_1 =& \frac{1}{\sqrt{2}}\log_{10}(R_\mathrm{e}\sigma\rs{c}^2)
&\kappa_2 =
\frac{1}{\sqrt{6}}\log_{10}(\frac{I_\mathrm{e}^2\sigma\rs{c}^2}{\re}) \nonumber\\
\kappa_3 =&
\frac{1}{\sqrt{3}}\log_{10}(\frac{\sigma\rs{c}^2}{I_\mathrm{e}R_\mathrm{e}})&
\end{eqnarray}
$I\rs{e}$ is the effective surface brightness or the surface
brightness within a distance \re from the galaxy centre. The physical
meaning of the $\kappa$ variables can be described as follows:
$\kappa_1\propto \log_{10}(M), \kappa_2\propto \log_{10}(I\rs{e}^3
M/L), \kappa_3\propto \log_{10}(M/L)$. Our models are plotted in
$\kappa$-space in Fig. \ref{FPkappa}. We show an edge-on projection
of the FP in panel a), a face-on projection in panel b). In Fig.
\ref{FP2_1} we plot the FP using observed quantities
($R\rs{e},\sigma\rs{c},I\rs{e})$ as coordinates. Fig. \ref{FP2_2}
shows the deviation of galaxies from the FP. The observational data on
panel a) of Fig. \ref{FPkappa} learns that the trend of decreasing
$M/L$ with decreasing mass is reversed at $\kappa_1\approx 2.5$. This
indicates that dwarf galaxies and subsequent lighter systems suffer
from depletion of baryonic matter relative to dark matter. Our models
clearly reproduce this trend. From panel b) we learn that, despite a
higher mass-to-light ratio, $\kappa_2$ is lower for DGs and
light-weight galaxies than for more massive galaxies on the FP. These
light-weight galaxies are therefore much fainter and more diffuse than
their FP counterparts. Again we clearly reproduce this trend with our
models. Figs. \ref{FP2_1} and \ref{FP2_2} confirm this picture~: our
models reproduce the deviation of dSphs and dEs from the FP quite
well. This deviation can hence be entirely contributed to the
shallowness of the galaxy potential well, no stripping effects need to
be invoked.
%
%
%
\section{Discussion and conclusions}
\label{sec:sum}
We have presented N--body/SPH models of dwarf galaxies. The basic SPH
scheme was modified to include star formation, supernova feedback,
chemical enrichment and gas cooling. The simulations start from a
homogeneous gas cloud collapsing onto a DM halo. A first series of
models (A models) was used to investigate the influence of the gas
particle number on the overall evolution of the system. We found that
a minimum of 25$\,$000 particles is required to resolve the evolution
accurately. Using less particles leads to a reduced first star
formation peak. Because of this, the ISM is injected with less energy
(less SNe), resulting in higher subsequent SF peaks and hence a larger
final stellar mass (Table \ref{tabnpart}). Based on these experiments,
we decided to use 30$\,$000 gas particles and 15$\,$000 DM particles
in the simulations.

All our model galaxies exhibit burst-like star formation. This is a
direct consequence of star formation in a shallow potential well.
Supernovae heat the gas, causing it to expand. The gas density
consequently drops below the threshold $\rho\rs{c}$ for star
formation. After the lapse of an adiabatic cooling period, the gas is
again allowed to cool and contract. Star formation thus appears to be
a self-regulating mechanism. 
Since substantial amounts of gas remain in and around the model
galaxies throughout the duration of the simulations (see
fig. \ref{cummassfig}), one expects that tidal interactions and
ram-pressure stripping in dense environments will have a significant 
effect on their SFHs.

We compared the model galaxies with observational data in terms of
morphology (S\'ersic $n$, $R\rs{e}$, $\mu_0$), kinematics
($\sigma\rs{1D,0}$), colour, metallicity and location with respect to
the FP ($\kappa_1,\kappa_2,\kappa_3$). Despite their simplicity, the
models do very well in the range $-8 > M\rs{B} > -15$ mag. The models
reproduce the slopes of the size, colour, metallicity, and velocity
dispersion vs. luminosity relations very well. The zero-points of these
relations are reproduced to within a factor of 2. The position of the
model dwarf galaxies in the three-dimensional space spanned by
$(R\rs{e},\sigma\rs{1D,0},I\rs{e})$ agrees excellently with the
observations. Dwarf galaxies trace a sequence within the FP (i.e. in
the $(\kappa_1,\kappa_2)$ projection) that is almost orthogonal to
that of massive galaxies. In the dwarf regime, the height above the FP
(in the $(\kappa_1,\kappa_3)$ projection) is a decreasing function of
galaxy mass. This is a consequence of less massive galaxies having
shallower potentials than more massive ones, decreasing the tendency
of the gas to fall into the potential well and increasing the
capability of supernovae to support the gas by heating it.

Furthermore we found that a good check for the validity of models is
the combination of \re and $\sigma\rs{c}$. \re is very sensitive to
the location of star formation and hence to the time evolution of the
density of the ISM.  $\sigma\rs{c}$ reflects, through the virial
theorem, the total amount and distribution of (dark) matter. We
demonstrate this using our B and C models, where it can be clearly
seen (figs. \ref{MB_re} and \ref{MB_disp}) that using a more diffuse
DM halo indeed raises $R\rs{e}$, and at the same time reduces
$\sigma\rs{c}$.
\section*{Acknowledgments}
We wish to thank the anonymous referee for valuable remarks that very
much improved the content and presentation of the paper. SV thanks
Johan Maes for a critical reading of the original version of the paper
and several helpful suggestions. SV acknowledges and is grateful for
the financial support of the Fund for Scientific Research -- Flanders
(FWO). Part of the simulations were run on our local computer cluster
ITHILDIN.
\bibliographystyle{mn2e} 
\bibliography{biblio}
\label{lastpage}
\end{document}